\documentclass[aps,pra,superscriptaddress,longbibliography,nofootinbib,twocolumn]{revtex4-2}

\usepackage{graphicx}
\usepackage{enumitem}
\usepackage{amsmath}
\usepackage{amssymb}
\usepackage{hyperref}
\usepackage[utf8]{inputenc}
\usepackage{mathtools}
\usepackage[english]{babel}
\usepackage{bbm}
\hypersetup{colorlinks=true, linkcolor=blue, citecolor=blue, urlcolor=blue}
\usepackage{xcolor}
\usepackage{braket}
\usepackage[normalem]{ulem}
\DeclareMathAlphabet{\mathbbold}{U}{bbold}{m}{n}

\newcommand{\ie}{i.\,e.,\ }

\newcommand{\eg}{e.\,g.,\ }

\newcommand{\ii}{\mathrm{i}}

\newcommand{\Tr}{\operatorname{Tr}}
\newcommand{\rr}{\mathbf{r}}
\newcommand{\pp}{\mathbf{p}}

\newcommand{\ketbra}[2]{\left|{#1}\middle\rangle\middle\langle{#2}\right|}
\newcommand{\pa}[1]{\partial_{#1}}
\newcommand{\rhop}{\hat{\rho}}
\newcommand{\Hop}{\hat{H}}
\newcommand{\pare}[1]{\left( {#1} \right)}
\newcommand{\spare}[1]{\left[ {#1} \right]}

\newcommand{\rrop}{\hat{\rr}}

\newcommand{\eqnref}[1]{Eq.~\eqref{#1}}
\newcommand{\figref}[1]{Fig.~\ref{#1}}

\newcommand{\appref}[1]{Appendix~\ref{#1}}
\newcommand{\w}{\omega}

\newcommand{\id}{\mathbbold{1}}
\def\dpp{\boldsymbol{\wp}}
\newcommand{\W}{\Omega}
\newcommand{\uu}{\mathbf{u}}
\newcommand{\bop}{\hat{b}}
\newcommand{\bdop}{\hat{b}^\dag}
\newcommand{\Rop}{\hat{R}}

\newcommand{\hc}{\text{h.c.}}

\newcommand{\eref}[1]{\text{Eq.}~\eqref{#1}}

\newcommand{\muop}{\hat{\mu}}

\newcommand{\avg}[1]{\ensuremath{\langle #1 \rangle}}

\begin{document}
\title{Collectively enhanced ground-state cooling in subwavelength atomic arrays}

\author{Oriol Rubies-Bigorda}
\email{orubies@mit.edu}
\affiliation{Physics Department, Massachusetts Institute of Technology, Cambridge, Massachusetts 02139, USA}
\affiliation{Department of Physics, Harvard University, Cambridge, Massachusetts 02138, USA}
\author{Raphael Holzinger}
\affiliation{Department of Physics, Harvard University, Cambridge, Massachusetts 02138, USA}
\affiliation{Institute for Theoretical Physics, University of Innsbruck, A-6020 Innsbruck, Austria}
\author{Ana Asenjo-Garcia}
\affiliation{Department of Physics, Columbia University, New York, New York 10027, USA}
\author{Oriol Romero-Isart}
\affiliation{ICFO - Institut de Ciencies Fotoniques, The Barcelona Institute of Science and Technology, 08860 Castelldefels (Barcelona), Spain}
\affiliation{ICREA - Institucio Catalana de Recerca i Estudis Avançats, 08010 Barcelona, Spain}
\author{Helmut Ritsch}
\affiliation{Institute for Theoretical Physics, University of Innsbruck, A-6020 Innsbruck, Austria}
\author{Stefan Ostermann}
\affiliation{Department of Physics, Harvard University, Cambridge, Massachusetts 02138, USA}
\author{Carlos Gonzalez-Ballestero }
\affiliation{Institute for Theoretical Physics, Vienna University of Technology (TU Wien), 1040
Vienna, Austria}
\author{Susanne F. Yelin}
\affiliation{Department of Physics, Harvard University, Cambridge, Massachusetts 02138, USA}
\author{Cosimo C. Rusconi }
\email{cr3327@columbia.edu}
\affiliation{Department of Physics, Columbia University, New York, New York 10027, USA}
\affiliation{Instituto de F\'isica Fundamental - Consejo Superior de Investigaciones Cient\'ifica (CSIC), Madrid, Espa\~na}

\begin{abstract}
Subwavelength atomic arrays feature strong light-induced dipole-dipole interactions, resulting in subradiant collective resonances characterized by narrowed linewidths. In this work, we present a sideband cooling scheme for atoms trapped in subwavelength arrays that utilizes these narrow collective resonances. Working in the Lamb-Dicke regime, we derive an effective master equation for the atomic motion by adiabatically eliminating the internal degrees of freedom of the atoms, and validate its prediction with numerical simulations of the full system. Our results demonstrate that subradiant resonances enable the cooling of ensembles of atoms to temperatures lower than those achievable without dipole interactions, provided the atoms have different trap frequencies. Remarkably, narrow collective resonances can be sideband-resolved even when the individual atomic transition is not. In such scenarios, ground-state cooling becomes feasible solely due to light-induced dipole-dipole interactions. This approach could be utilized for future quantum technologies based on dense ensembles of emitters, and paves the way towards harnessing many-body cooperative decay for enhanced motional control.
\end{abstract}

\maketitle

\section{Introduction}

Laser cooling of atoms is a cornerstone of atomic physics and quantum technologies, enabling applications such as high-precision spectroscopy and metrology~\cite{wineland1979laser,hansch2006nobel}. 
This technique relies on enhancing light scattering processes that dissipate atomic kinetic energy, thereby cooling their motion.
In free space, harmonically trapped atoms can be cooled to their motional ground state via sideband cooling~\cite{Neuhauser1978}, where the laser is tuned to a red-detuned motional sideband of an atomic transition.
This ensures that each absorption and subsequent emission of a photon reduces the atom’s motional energy.
Sideband cooling is most effective when the atoms are tightly confined, such that the trap frequency greatly exceeds the natural linewidth of the transition. When a sufficiently narrow transition is unavailable, alternative methods can be used to engineer favorable scattering dynamics.
Notable examples include Raman sideband cooling, which employs Raman transitions to realize effective narrow-linewidth transitions between metastable states~\cite{Marzoli1994,Vuletic1998,Hamann1998,Perrin1998}, and cavity cooling, where an optical resonator modifies the atom’s scattering cross section~\cite{Cirac1995,Horak1997,Vuletic2000,Vuletic2001,Domokos2001,Zippilli2005PRL,Zippilli2005PRA}.

\begin{figure}[h!]
    \centering
    \includegraphics[width=\columnwidth]{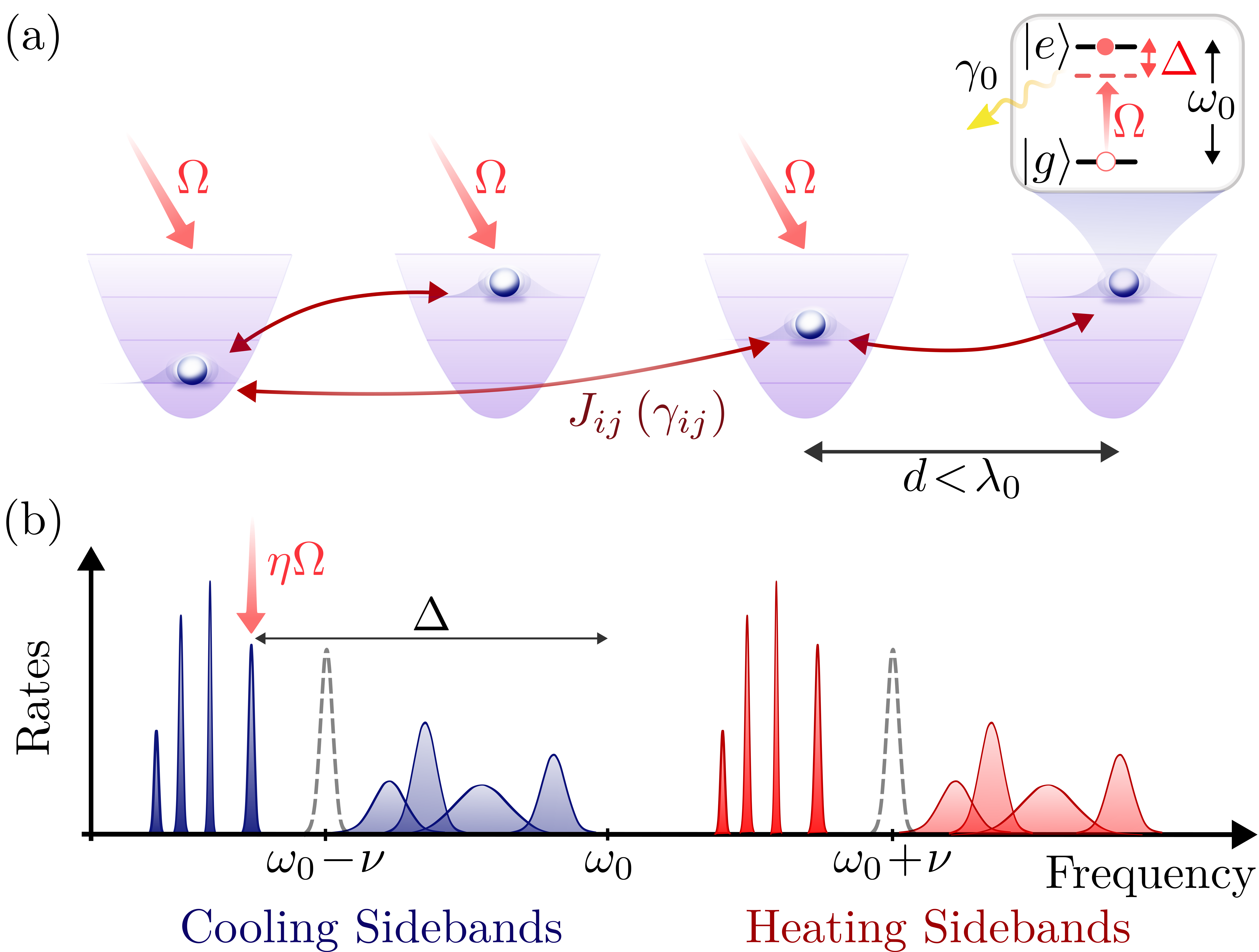}
    \caption{Motional cooling of a subwavelength array of trapped atoms. (a) Dipole-interacting two-level atoms driven by an external laser. Collective narrowed and broadened resonances emerge from the long-range coherent (dissipative) dipole-dipole interactions $J_{ij}$ ($\gamma_{ij}$). (b) Motional sidebands of the collective resonances of the subwavelength array. Narrow sidebands, associated to subradiant collective resonances, can be resonantly driven (red arrow) to cool atoms to a lower temperature. Sidebands of a single atoms are indicated by the gray dashed line.}
    \label{fig:Schematic}
\end{figure}

The scattering properties of atoms in free space can be profoundly altered when the atoms are arranged at separations comparable to or smaller than the wavelength of the relevant dipole transition. In this regime, light-induced dipole-dipole interactions give rise to a strong collective optical response characterized by broad (superradiant) and narrow (subradiant) resonances~\cite{AsenjoGarcia2017,Bettles2016}. This collective response, which has recently been observed in ordered atomic arrays~\cite{Rui2020,Srakaew2023},  can be harnessed to enhance the sensitivity of atomic lattice clocks~\cite{Henriet2019,Hutson2024}, improve the fidelity of quantum memories~\cite{AsenjoGarcia2017}, create pristine all-atomic material to control light propagation~\cite{Shahmoon2017,Masson2020,Bekenstein2020,Brechtelsbauer2021,CastellsGraells2024}, and engineer controlled interaction between quantum emitters~\cite{CastellsGraells2021,RubiesBigorda2022PRR,BuckleyBonanno2022}. Crucially, maintaining an ordered lattice of cold atoms is essential for these applications.
It is thus natural to ask what is the impact of dipolar interactions on laser cooling of atoms in a subwavelength array, and whether their collective optical response can be used to improve the cooling performances without resorting techniques such as Raman sideband cooling or cavity cooling.

Previous studies have addressed this problem leading to different and often contradicting results.
The case of two ions was studied by Vogt et al.~\cite{Vogt1996} unveiling regimes of enhanced cooling due to both subradiance and interference between different scattering processes. It is however unclear whether this mechanism generalizes to many atoms.
For large one-dimensional atomic arrays, Palmer and Beige~\cite{Palmer2010} demonstrated that dipole interaction can lead to a $30\%$ reduction of the final motional temperature. The scope of their work is however limited to a minimum interatomic separation because their description breaks down in the presence of perfectly dark states.
Smaller separations were considered by
Shahmoon et al.~\cite{Shahmoon2019,Shahmoon2020} and Robicheaux and co-workers~\cite{Robicheaux2019,Suresh2021}, demonstrating that collective dipole interactions generally lead to excess motional heating in two-dimensional (2D) arrays when the timescale of the motion is respectively much longer and much shorter than the scattering dynamics. 
To further complicate the picture, Wang et al~\cite{Wang2023} demonstrated that a single target emitter in free space can be cooled to a slightly lower temperature if it is dipole coupled to a small surrounding ring of emitters, albeit only for specific configurations that reduce coherent-dipole exchange between atoms. However, this strategy cannot be directly generalized to enhance the cooling of multiple emitters simultaneously.

The present article addresses this outstanding controversy through accurate analytical and numerical modeling of the cooling mechanisms. The work is based on the derivation of an effective master equation for the atomic motional dynamics for large subwavelength arrays. Our effective model overcomes the limitations of previous approaches~\cite{Palmer2010} by recognizing that subradiant collective linewidths are broadened by atomic center-of-mass (vacuum and thermal) fluctuations~\cite{Guimond2019,Rusconi2021}.
We identify four key ingredients that enable ground state cooling of atoms in subwavelength arrays [Fig.~\ref{fig:Schematic}.(a)], irrespective of the array geometry and across a broad range of interparticle distances. 
(1) As the final motional temperature for sideband cooled atoms is set by the transition linewidth, narrow collective sidebands can be addressed to reach lower final temperatures.
(2) \emph{Motional sidebands} of collective subradiant states can be excited using simple driving configurations, such as a uniform drive that addresses a bright carrier transition [Fig.~\ref{fig:Schematic}.(b)].
(3) Dipole-mediated interactions facilitate motional excitation exchange between atoms, which can increase the final temperature despite the reduced linewidths. This detrimental excitation exchange can be suppressed by modifying the trap frequencies of different atoms, rendering the interactions effectively off-resonant.
(4) The main impairment to cooling comes from the coalescence between the chosen narrow cooling sideband and the heating sideband of another collective mode (either subradiant or superradiant). 
By combining these four insights, we identify the parameter regimes where subwavelength arrays of up to a few tens of atoms can be cooled to temperatures below those achievable in the absence of interactions. Remarkably,  we find conditions where ground-state cooling is only possible via collective dipole-dipole interactions.

The remainder of the article is structured as follows. In Sec.~\ref{Section: System_Model}, we introduce the model and derive the effective master equation for the atomic motion. After reviewing the cooling of independent (\ie non-interacting) atoms in Sec.~\ref{sec:independent}, we present, in Sec.~\ref{Section: cooling_2_atoms}, the first main results of our work: the collective cooling of two dipole-interacting atoms. This simple example illustrates the main conceptual contributions of this study. We then generalize these results to arrays of $N$ atoms in Sec.~\ref{Section: cooling_N_atoms}. Our conclusions are summarized in Sec.~\ref{sec:Conclusions}. The article is complemented by several appendixes; in particular, Appendix~\ref{app:Derivation_effective_ME} details the derivation of the effective master equation for atomic motion, which incorporates collective effects.

\section{System and Model}
\label{Section: System_Model}

We consider an ensemble of $N$ two-level atoms with mass $M$ and resonant dipole transition ($\ket{g}\leftrightarrow \ket{e}$) at a frequency $\w_0= c k_0 = 2\pi c / \lambda_0$ and linewidth $\gamma_0$.
The atoms are harmonically trapped at positions $\mathbf{r}_{0j}$ and are driven by an external monochromatic laser at frequency $\omega_L = \omega_0 + \Delta$ [Fig.~\ref{fig:Schematic}(a)]. 
Their internal degrees of freedom are described by the lowering operators $\hat{\sigma}_j\equiv\ket{g_j}\bra{e_j}$, while their motional degrees of freedom are described by the center-of-mass momentum $\hat{\pp}_j$ and position $\hat{\mathbf{r}}_j = \mathbf{r}_{0j} + \delta \hat{\mathbf{r}}_j$ operators, where $\delta \hat{\mathbf{r}}_j$ denotes the displacement from the trap center. Following this description, we will refer to internal excitation of the atoms as spin excitations.
Throughout this work, we focus on ordered atomic arrays with subwavelength interatomic spacing $d<\lambda_0$. Nevertheless the results of this section apply more broadly to ensembles of trapped atoms with arbitrary geometries and interatomic separations.

The effective dynamics of the emitters can be derived from the total Hamiltonian describing their interaction with the electromagnetic field within the dipole approximation. By tracing out the field degrees of freedom under the Born-Markov approximation, the evolution of the atomic state $\muop$ is governed by the master equation~\cite{Stenholm1986,Palmer2010}
\begin{equation}\label{eq:ME_Total}
	\pa{t}\muop = - \frac{i}{\hbar}\left[ (\Hop_\text{A} + \Hop_\text{L}) \muop-\muop (\Hop^\dag_\text{A} + \Hop_\text{L})\right] + \mathcal{D}\muop,
\end{equation}
The term in the square brackets describes the non-hermitian evolution of the system and contains two contributions. The first one describes the dynamics of the atomic array in the absence of a drive and reads
\begin{equation}\label{eq:H_A}
\begin{split}
    \Hop_\text{A} \equiv& \sum_j \spare{ \frac{\hat{\pp}^2_j}{2M} + \frac{M\nu_j^2}{2}  (\rrop_j-\rr_{0j})^2}- \hbar \Delta \sum_j \hat{\sigma}^\dag_j\hat{\sigma}_j\\
    &+ \hbar \sum_{i,j} G(\rrop_i-\rrop_j)\hat{\sigma}^\dag_i\hat{\sigma}_j, 
\end{split}
\end{equation}
where $\nu_{j}$ is the trap frequency of site $j$. The first two terms in \eqnref{eq:H_A} represent the mechanical and internal energy of the atoms, while the last term captures the non-Hermitian dipole-dipole interaction between the emitters, including recoil effects. This interaction is governed by
\begin{equation}\label{eq:G_real_imag}
	G(\rr) \equiv J(\rr)-\frac{\ii}{2}\gamma(\rr),
\end{equation}
where $J(\rr)$ and $\gamma(\rr)$ represent the coherent and dissipative contributions, respectively. At the origin, we define $G(\mathbf{0})=-i\gamma_0/2$. The explicit form of the dipole-coupling rate~\eqref{eq:G_real_imag} depends on the electromagnetic Green's tensor that describes how the photon propagates and mediate interactions between the atoms. In the following, we consider atoms in free space for which the explicit form of \eqnref{eq:G_real_imag} is given in  Appendix~\ref{app:Green_Tensor}.

The effect of the external laser drive is described by the Hamiltonian
\begin{equation}\label{eq:H_L}
    \hat{H}_L = \hbar\sum_j \spare{ \Omega (\rrop_j) \hat{\sigma}_j^\dagger + \Omega^* (\rrop_j) \hat{\sigma}_j},
\end{equation}
where $\W(\rrop_j)$ represents a possibly space-dependent Rabi frequency.

The last term in \eqnref{eq:ME_Total} is the population recycling term and reads
\begin{equation}
	\mathcal{D}\muop = \gamma_0\sum_{i,j}\int \!\! \text{d}\uu\,\mathcal{N}(\uu) e^{-i k_0 \uu\cdot \rrop_j}\hat{\sigma}_j \muop e^{i k_0 \uu\cdot \rrop_i}\hat{\sigma}_i^\dag,
\end{equation}
where $\uu$ is a unit vector and the integration is taken over all possible directions of $\uu$. $\mathcal{N}(\uu)$ is the dipole-emission pattern, which depends on the atomic polarization and is normalized such that $\int\!\text{d}\uu\,\mathcal{N}(\uu)=1$.

For clarity, we restrict the discussion below to atomic motion along a single spatial direction $\alpha=x,y,z$. More general expressions that account for motion along all three directions are provided in Appendix~\ref{app:Formalism_Multiple_Directions}.

\subsection{Lamb-Dicke approximation}\label{sec:Lamb-Dicke_approx}

We consider the system to be in the Lamb-Dicke regime, such that for all atoms 
\begin{equation}\label{eq:Lamb-Dicke}
	\eta_j (2\bar{n}_j + 1) = \sqrt{\frac{\nu_R}{\nu_j}}(2\bar{n}_j + 1)  \ll 1
\end{equation}
where $\eta_j = \sqrt{\nu_R/\nu_j}$ is the Lamb-Dicke parameter for atom $j$, $\bar{n}_j$ the its thermal average phonon population, and $\nu_R\equiv \hbar k_0^2/2M$ is the atomic recoil frequency.
When Eq.~\eqref{eq:Lamb-Dicke} holds, atomic motion is confined to a region much smaller than the resonant transition wavelength $\lambda_0$. We remark that for subwavelength atomic arrays this condition is automatically satisfied.
Fluctuations around the trapping sites can then be perturbatively treated by expanding Eq.~(\ref{eq:ME_Total}) up to second order in $\delta\hat{\rr}_j$ to obtain
\begin{equation}\label{eq:ME_Lamb-Dicke}
	\partial_t \muop = \pare{\mathcal{L}_0 + \eta \mathcal{L}^{(1)}_\text{int} + \eta^2 \mathcal{L}^{(2)}_\text{int}} \muop,
\end{equation}
where we have defined the \emph{averaged} Lamb-Dicke parameter $\eta \equiv \sqrt{ \nu_R / \bar{\nu}}$, where $\bar{\nu}\equiv \sum_j \nu_j/N$. This quantity characterizes the strength of the interactions between the internal and mechanical degrees of freedom of the atoms, provided that $\nu_i \approx \nu_j$ $\forall i,j$.

The first term in Eq.~(\ref{eq:ME_Lamb-Dicke}) describes the uncoupled dynamics of the internal and motional degrees of freedom and reads
\begin{equation}\label{eq:L_0}
\begin{split}
	\mathcal{L}_0 \muop =& - \frac{i}{\hbar}\pare{\Hop_0\muop - \muop \Hop_0^\dag}+ \sum_{i,j} \tilde{\gamma}_{ij} \hat{\sigma}_j \muop \hat{\sigma}_i^\dag,
\end{split}
\end{equation}
where $\tilde{\gamma}_{ij} \equiv \gamma_{ij}+(\eta_i^2 +\eta_j^2)(1-\delta_{ij}) \gamma''_{ij} / 2$. The non-Hermitian Hamiltonian $\hat{H}_0$ reads
\begin{equation}\label{eq:H_0}
    \hat{H}_0 = \hbar\sum_j \nu_j \bdop_j\bop_j + \hbar \sum_j \left( \Omega_j \hat{\sigma}_j^\dagger + \Omega_j^* \hat{\sigma}_j \right)+ \Hop_\text{S},
\end{equation}
where we defined the bosonic annihilation (creation) operator $\hat{b}_j$ ($\bdop_j$) which creates (destroys) a motional quantum (phonon) at site $j$. The first term is the mechanical energy of the atoms in the trap, while the second term describes the carrier excitation of the laser, $\ket{g_j}\leftrightarrow \ket{e_j}$, which does not change the phonon number. The last term reads
\begin{equation}\label{eq:H_S}
	\Hop_\text{S} = - \hbar\Delta\sum_j \hat{\sigma}^\dag_j\hat{\sigma}_j + \hbar \sum_{i,j}\tilde{G}_{ij} \hat{\sigma}^\dag_i\hat{\sigma}_j.
\end{equation}
Here, we defined $\tilde{G}_{ij} \equiv G_{ij} + (\eta_i^2 +\eta_j^2) (1-\delta_{ij})G''_{ij}/2$, where $G_{ij}\equiv G(\rr_{0i}-\rr_{0j})$ is the dipole-dipole coupling for atoms fixed at the trap centers and $G_{ij}''\equiv  k_0^{-2} \partial^2 G (\mathbf{r})/\partial^2 r_\alpha |_{\rr_{0i}-\rr_{0j}} \equiv J_{ij}'' - i \gamma_{ij}''/2$.
In agreement with our definition of $G(\mathbf{0})$, we have $G_{jj}''  = -i \gamma_{jj}''/2$.
We emphasize that \eqnref{eq:L_0} contains corrections to second order in the Lamb-Dicke parameter that appear inside the definition $\tilde{G}_{ij}$ and $\tilde{\gamma}_{ij}$. These terms arise from the second-order Lamb-Dicke expansion of \eqnref{eq:ME_Total}, as we discuss below. 

The second term in \eqnref{eq:ME_Lamb-Dicke} corresponds to the spin-motional coupling to first order in the Lamb-Dicke parameter. It reads
\begin{equation}\label{eq:L_1}
\begin{split}
	\mathcal{L}_\text{int}^{(1)} \muop =& \frac{-i}{\hbar}\!\pare{\!\Hop_1\muop \!-\! \muop\Hop_1^\dag}\!+\!\sum_{i\neq j}\! \gamma_{ij}' \hat{\sigma}_j\! \left( \muop  \hat{R}_i\! -\! \hat{R}_j  \muop \right)\! \hat{\sigma}_i^\dagger
\end{split}
\end{equation}
where we defined the dimensionless position operator $\hat{R}_j\equiv \sqrt{\bar{\nu}/\nu_j}(\bdop_j+\bop_j)$ and the non-Hermitian Hamiltonian is obtained from the first order Lamb-Dicke expansion of Eqs.~\eqref{eq:H_A} and \eqref{eq:H_L}. It reads
\begin{equation}\label{eq:H_1_Lamb-Dicke}
    \frac{\hat{H}_{1}}{\hbar}\! = \! \sum_j\! \left( \Omega_j'  \hat{\sigma}_j^\dagger\! +\hc \right)\!\!  \hat{R}_j\! +\! \sum_{i,j} \!G_{ij}' \!\!\left( \hat{R}_i\! -\! \hat{R}_j \right)\! \hat{\sigma}_i^\dagger \hat{\sigma}_j,
\end{equation}
where we defined $\W'_j \equiv k_0^{-1}\partial \W(\rr)/\partial r_{\alpha}|_{\rr_{0j}}$, and $G_{ij}' \equiv J_{ij}' - i \gamma_{ij}'/2 = k_0^{-1} \partial G (\mathbf{r})/\partial r_\alpha |_{\mathbf{r}_{0i}-\mathbf{r}_{0j}}$. Note that $G'_{jj}=0$.
The first term in \eqnref{eq:H_1_Lamb-Dicke} describes the sideband excitation induced by the laser drive, in which photon absorption leads to the excitation of an atom accompanied by the gain or loss of a motional quantum, $\ket{g_j,n_j} \leftrightarrow \ket{e_j,n_j\pm 1}$, where $n_j$ is the number of phonons at site $j$.
The second term in \eqnref{eq:H_1_Lamb-Dicke} describes instead the mechanical effects of dipole-dipole interaction. In this case, a spin excitation is transferred between atoms $i$ and $j$ by simultaneously creating or annihilating a phonon at either site,
$\ket{e_i,g_j}\ket{n_i,n_j}\leftrightarrow \ket{g_i,e_j}\ket{n_i\pm 1,n_j}$ or $\ket{e_i,g_j}\ket{n_i,n_j}\leftrightarrow \ket{g_i,e_j}\ket{n_i,n_j\pm 1}$. 

The last term in \eqnref{eq:ME_Lamb-Dicke} contains interaction terms to second order in the Lamb-Dicke parameter. It reads
\begin{equation}\label{eq:L_2}
\begin{split}
	\mathcal{L}_\text{int}^{(2)}\muop =& - \frac{i}{\hbar}\pare{\Hop_2 \muop - \muop \Hop_2^\dag} - \sum_{i,j}\gamma_{ij}''\hat{\sigma}_j \Rop_j \muop \Rop_i\hat{\sigma}_i^\dag  \\
	+&  \sum_{i, j} \frac{\gamma_{ij}''}{2} \hat{\sigma}_j \spare{ \muop\Rop_i^2 \! + \!  \hat{R}_j^2 \muop - \frac{\eta_i^2+\eta_j^2}{\eta^2}\muop} \hat{\sigma}_i^\dagger.
\end{split}
\end{equation}
Here we defined the nonhermitian Hamiltonian
\begin{equation}\label{eq:H_2}
\begin{split}
	\frac{\Hop_2}{ \hbar} &= \sum_j \left( \frac{\Omega_j''}{2}  \hat{\sigma}_j^\dagger + \hc \right)  \hat{R}_j^2 \\ &+\sum_{i,j} \frac{G_{ij}''}{2} \left[ \big(\hat{R}_i - \hat{R}_j\big)^2 - \frac{\eta_i^2 + \eta_j^2}{\eta^2} \right] \hat{\sigma}_i^\dagger \hat{\sigma}_j,
\end{split}
\end{equation}
where $\Omega_j'' \equiv k_0^{-2} \partial^2 \Omega (\mathbf{r})/\partial^2 r_\alpha |_{\mathbf{r}_{0j}}$, and $G''_{ij}$ is defined above. The population recycling terms, respectively the second and last term in \eqnref{eq:L_2}, capture the diffusion of the center-of-mass motion induced by photon scattering, now occurring in a correlated fashion due to dipole-dipole interactions.
Equations \eqref{eq:L_2} and \eqref{eq:H_2} are obtained from the second order Lamb-Dicke term of \eqnref{eq:ME_Total}, except for the term that, as a result of normal ordering of the phonon operators, is independent of the motional degrees of freedom. 
We have subtracted this contribution from \eqref{eq:L_2} and \eqref{eq:H_2} and included it in  \eqnref{eq:L_0}. It describes effects of the finite spatial width of the ground-state wavefunctions in the harmonic traps, that effectively smears out the dipole-dipole interactions over the extent of the atomic motion. 
This modifies the linear optical response of the system as compared with the case of nonmoving atoms, as we shall now illustrate.

The system's linear optical response is obtained from the single-excitation spectrum of \eqnref{eq:H_S}. We define the eigenstates $\ket{\phi_\lambda}= \sum_j \phi_{\lambda,j} \hat{\sigma}_j^\dagger\ket{g}$ such that $H_S\ket{\phi_\lambda} = \epsilon_\lambda \ket{\phi_\lambda}$. The complex eigenvalues $\epsilon_\lambda = -\Delta + J_\lambda - i \gamma_\lambda /2$ describe the collective energy shifts $J_\lambda$ and collective linewidth $\gamma_\lambda$ of the transition $\ket{g}^{\otimes N}\leftrightarrow \ket{\phi_\lambda}$. 
As a consequence of interference between the field emitted by nearby atoms in the array, the linewedith $\gamma_\lambda$ is strongly modified leading to bright ($\gamma_\lambda\gg\gamma_0$) and dark ($\gamma_\lambda\ll\gamma_0$) collective resonances.
While for an array of pinned atoms, collective dark states have a vanishing linewidth in the limit of $N\rightarrow\infty$, this is not the case for \eqnref{eq:H_S}. The finite extent of the atomic ground state motional wavefunction broaden the collective linewdith such that~\cite{Guimond2019,Rusconi2021}
\begin{equation}\label{eq:gamma_sub_lower_bound}
	\gamma_\text{min} \equiv \min_{\lambda}(\gamma_\lambda) \gtrsim \eta^2\gamma_0.
\end{equation}
This key feature prevents the emergence of perfectly dark states. We remark that the broadening in \eqnref{eq:gamma_sub_lower_bound} is solely due to vacuum fluctuations.  
Additional corrections arise from finite temperature contribution and from dipole-induced forces. These terms, respectively included in Eqs.~\eqref{eq:L_1} and \eqref{eq:L_2}, are of the same order in $\eta^2$ and contribute to the broadening of the collective linewidth (see \appref{app:Derivation_effective_ME}).

The master equation in the Lamb-Dicke approximation, \eqnref{eq:ME_Lamb-Dicke}, can be simulated to study the coupled spin and motional dynamics of systems with a few emitters and a small number of motional excitations (Appendix~\ref{app:Numerics}). 
To gain deeper insight into the cooling dynamics, as well as to make prediction for large system sizes, we derive an effective model for the motion of the atoms in the next subsection. 

\subsection{Effective master equation for the motional dynamics}
\label{subsection: Formalism_Effective_ME}

We derive an effective master equation for the atomic motion by adiabatically eliminating the spin degrees of freedom. Specifically, we make the following assumptions: (i) the system is weakly driven, such that its optical response remains in the linear regime; and (ii) the optomechanical coupling is weak, meaning that the coupling rates in \eqnref{eq:L_1} are smaller than the narrowest collective linewidth. For sideband excitations induced by the laser as given by \eqnref{eq:H_1_Lamb-Dicke}, this condition becomes $\eta |\W| \ll \gamma_\text{min}$. Importantly, the fact that dark states acquire a finite linewidth due to atomic position fluctuations—as captured by Eq.~\eqref{eq:gamma_sub_lower_bound}—is crucial for enabling the adiabatic elimination of subradiant spin modes. 
Hence, our effective model is valid also in the strongly subwavelength regime, $d\ll \lambda_0$, that couldn't be captured by the effective master equation derived in Ref.~\cite{Palmer2010}.

Treating the optomechanical interactions in \eqnref{eq:ME_Lamb-Dicke} perturbatively, we derive an effective master equation for the motional state $\hat{\rho}$ of the atoms up to order $O(\eta^2)$. In the frame rotating at the trap frequencies, it reads (see Appendix~\ref{app:Derivation_effective_ME})
\begin{equation}
\label{eq:ME_Motion}
\begin{split}
    \frac{d}{dt} \hat{\rho} &= -i\Big[\sum_{i,j=1}^N V_{ij}e^{\ii (\nu_i - \nu_j) t}\bdop_i\bop_j,\hat{\rho}\Big]\\
    &+ \sum_{i,j=1}^N\!\! \mathcal{R}_{ij}^-e^{\ii (\nu_i - \nu_j) t} \left( \hat{b}_j \hat{\rho} \hat{b}_i^\dagger- \frac{1}{2} \{\hat{b}_i^\dagger \hat{b}_j ,\hat{\rho}\} \right) \\
    &+ \sum_{i,j=1}^N\!\! \mathcal{R}_{ij}^+e^{-\ii (\nu_i - \nu_j) t} \left( \hat{b}_j^\dagger \hat{\rho} \hat{b}_i - \frac{1}{2} \{\hat{b}_i\hat{b}_j^\dagger  ,\hat{\rho}\} \right).
\end{split}
\end{equation} 
The first term describes a renormalization of the trap frequency ($V_{jj}$) and coherent exchange of phonons between different atoms ($V_{ij}$). The second and third terms describe collective cooling and heating processes, \ie the correlated loss and gain of a phonons or motional quanta among multiple atoms.
The coherent and dissipative couplings between different sites in \eqnref{eq:ME_Motion} arise from the delocalize nature of the collective response of the system, and it is a consequence of dipole interactions mediated by the vacuum electromagnetic field.

\begin{figure}
    \centering
    \includegraphics[width=0.9\columnwidth]{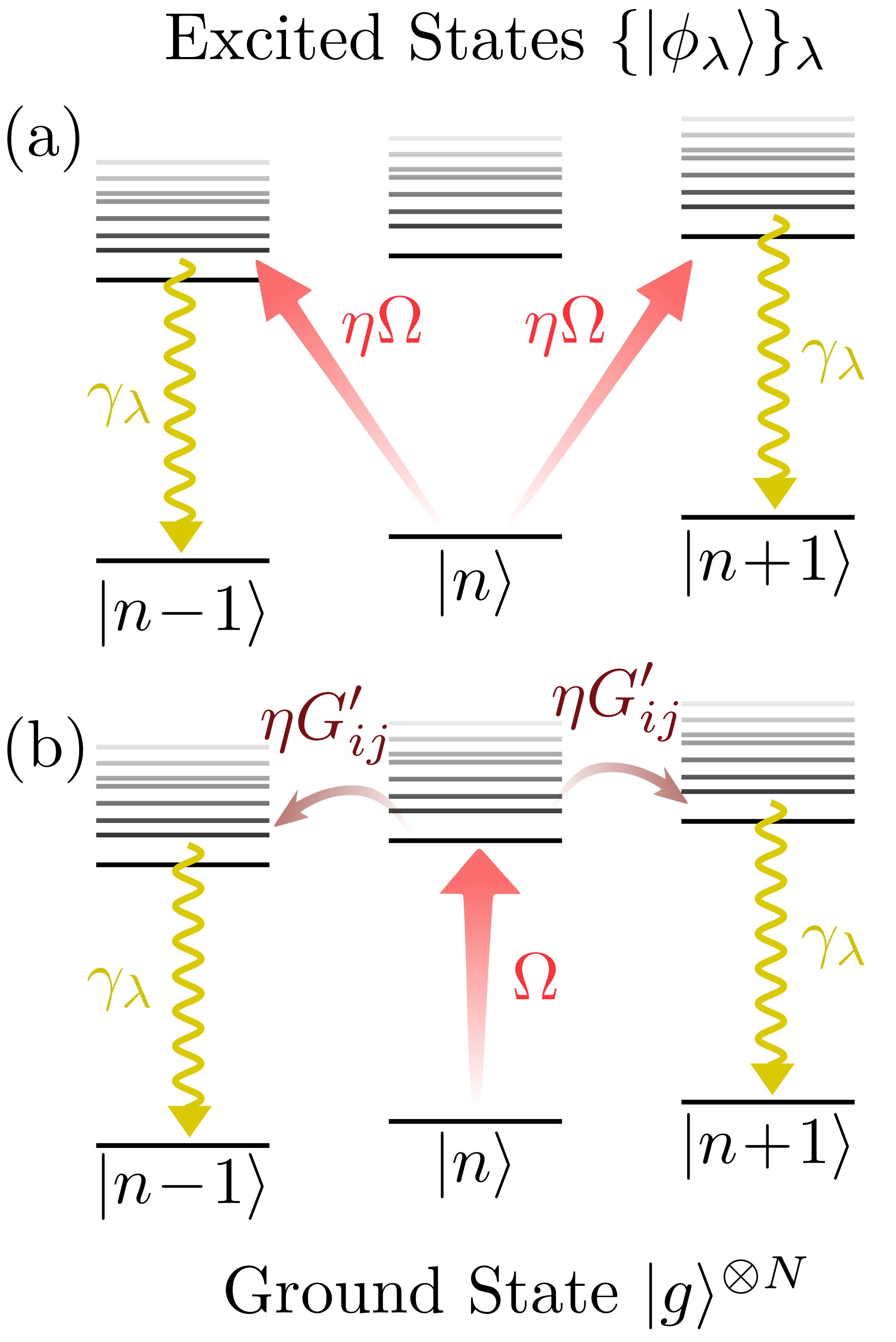}
    \caption{Scattering processes leading to cooling and heating. (a) Excitation of a collective sideband $\ket{g}^{\otimes N}\ket{n}\rightarrow \ket{\phi_\lambda}\ket{n\pm 1}$ by the laser followed by the collective decay to the ground state. The total number of phonons is labeled by $n$. (b) Excitation of a carrier transition $\ket{g}^{\otimes N}\ket{n}\rightarrow \ket{\phi_{\lambda'}}\ket{n}$. The excited state scatters to another excited state by either emitting or absorbing a phonon $\ket{\phi_{\lambda'}}\ket{n}\rightarrow \ket{\phi_{\lambda}}\ket{n\pm 1}$. Finally, the state collectively decays back to the ground state.}
    \label{fig:Cooling_Processes}
\end{figure}

The cooling and heating transition rates $\mathcal{R}_{ij}^\pm$ contain two main contributions. One arises from the second-order dipole-dipole interaction described in Eq.~\eqref{eq:L_2}, and corresponds to recoil-induced diffusion rate due to collective photon emission.
The other one is the result of interference between two distinct physical processes. 
The first process arises from the first-order spin-motion coupling induced by the drive [the term proportional to $\Omega_j'$ in Eq.~\eqref{eq:H_1_Lamb-Dicke}], where a ground-state atom absorbs a photon while simultaneously gaining or losing a phonon [see Fig.~\ref{fig:Cooling_Processes}(a)]. The subsequent spontaneous emission of the photon results in a net change of motional energy. 
The second process stems from the first-order spin-motion coupling induced by the dipole-dipole interactions, \ie the term proportional to $G_{ij}'$ in Eq.~\eqref{eq:H_1_Lamb-Dicke}. Here, the drive excites one collective state, that transitions to another by simultaneously creating or destroying a phonon [see Fig.~\ref{fig:Cooling_Processes}(b)]. Subsequent photon emission again results in the net loss or gain of a phonon. As a result the general formula for $\mathcal{R}_{ij}^{\pm}$ is rather lengthy and is given in Appendix~\ref{app:Derivation_effective_ME}. For simplicity, we consider here the case of atom pairs moving along a direction perpendicular to the line connecting their equilibrium positions. In this case, $G_{ij}' = 0$ and 
\begin{align}
\label{eq: dissipative_interactions_perp}
\mathcal{R}_{ij}^{\pm} = &-i \eta_i \eta_j \Omega_j' \Omega_i'^{*} \sum_{\lambda=1}^N \left(  \frac{c_{\lambda,ij}}{\epsilon_\lambda \pm \nu_{j}}  - \frac{c_{\lambda,ij}^*}{\epsilon_\lambda^* \pm \nu_{i}} \right) \nonumber \\ 
&-  \eta_i \eta_j \gamma_{ij}'' \avg{\sigma_j}_\text{ss}\avg{\sigma_i^\dag}_\text{ss}. 
\end{align}
The first term corresponds to cooling and heating due to the laser-induced spin-motion coupling. The contribution of each collective spin state $|\phi_\lambda\rangle$ is weighted by its overlap with atoms $i$ and $j$, quantified by $c_{\lambda,ij} = \phi_{\lambda,i} \phi_{\lambda,j} / \sum_p \phi_{\lambda,p}^2$. The second term describes recoil-induced diffusion, and is proportional to the laser-induced dipole moments of $i$ and $j$ atoms through the steady state coherences $\avg{\sigma_j}_\text{ss}$ and $\avg{\sigma_i^\dag}_\text{ss}$. 

The effective master equation in \eqnref{eq:ME_Motion} is gaussian and thus can be efficiently solved to compute the dynamics of the system. Furthermore, in certain regimes, the steady-state phonon population admits simple analytical expressions. In the following, we use \eqnref{eq:ME_Motion} to compute the cooling dynamics of dipole interacting atoms in different scenarios and benchmark these results with numerical simulation of the Lamb-Dicke master equation in \eqref{eq:ME_Lamb-Dicke}.

\section{Sideband cooling of independent atoms}\label{sec:independent}

We begin by reviewing cooling of independent atoms that do not interact via the common radiation field~\cite{Stenholm1986}, which naturally applies when interatomic distances are much larger than the transition wavelength ($d\gg \lambda_0$). 
In this regime, all couplings between the motional degrees of freedom vanish and Eq.~(\ref{eq:ME_Motion}) reduces to the well-known master equation for laser sideband cooling of individual atoms~\cite{cirac1992laser,eschner2003laser}.
For atoms in identical traps and driven uniformly, the dynamics of each atom are identical. Dropping the index $j$, the steady-state phonon number of each atom reads~\cite{cirac1992laser}
\begin{equation}\label{eq:n_ind}
    \bar{n}_{\mathrm{ind}} = \langle \hat{b}^\dagger \hat{b} \rangle = \frac{\mathcal{R}^+}{\mathcal{R}^- - \mathcal{R}^+},
\end{equation}
and the rate at which the phonon population decays over time is given by the cooling rate $\Gamma_\text{ind} \equiv \mathcal{R}^- - \mathcal{R}^+$, where the cooling ($\mathcal{R}^-$) and heating ($\mathcal{R}^+$) transition rates are
\begin{equation}
\label{eq:cool_heat_rates_independent}
    \mathcal{R}^{\pm} = \frac{\eta^2 |\Omega'|^2\gamma_0}{(-\Delta \pm \nu)^2 + (\gamma_0/2)^2}  + \frac{\eta^2 |\gamma_0''||\Omega|^2}{\Delta^2 + (\gamma_0/2)^2}.
\end{equation}
The transition rates $\mathcal{R}^{\pm}$ are the result of different physical processes whose contribution to Eq.~(\ref{eq:cool_heat_rates_independent}) can be tuned by adjusting the detuning $\Delta$. The first term of the cooling transition rate $\mathcal{R}^-$ corresponds to inelastic scattering of a laser photon accompanied by the annihilation of a phonon. This process, known as the red (or cooling) sideband, is resonant at a detuning $\Delta = -\nu$. Conversely, the first term of the heating transition rate $\mathcal{R}^+$ corresponds to inelastic scattering involving the creation of a phonon, and is referred to as the blue (or heating) sideband, centered at $\Delta = \nu$. Both sidebands are depicted as gray dashed Lorentzian line shapes in Fig.~\ref{fig:sketch}(b). The second term in Eq.~(\ref{eq:cool_heat_rates_independent}) is a Lorentzian of width $\gamma_0$ centered at the carrier transition, $\Delta = 0$, and describes difussion or recoil heating from spontaneous emission~\cite{cirac1992laser}.

For a plane-wave or Gaussian drive satisfying $\Omega' = i \Omega$, the steady-state phonon number is minimized at the optimal detuning $\Delta = -\sqrt{\nu^2 + (\gamma_0/2)^2}$ and is set by the ratio $\nu/\gamma_0$~\cite{Cohen1990LesHouces,eschner2003laser}. Ground-state cooling occurs in the resolved sideband regime, $\nu \gg \gamma_0$, where both the heating sideband and the carrier transition are far off-resonant.
In this limit, the optimal steady state phonon number is obtained for $\Delta \approx -\nu$. To leading order in the small parameter $\gamma_0/\nu$, it reads
\begin{equation}\label{eq:n_ind_sc}
    \bar{n}^{(\text{sc})}_\text{ind} \approx \frac{\gamma_0^2}{16 \nu^2}\pare{1 + 4  \frac{|\gamma_0^{''}|}{\gamma_0}} \approx \frac{13 \gamma_0^2}{80 \nu^2} \ll 1,
\end{equation}
and is reached at a rate $\Gamma_{\text{ind}}^{(sc)} \approx 4 \eta^2 \Omega^2 / \gamma_0$. Here, we have used the fact that $\gamma_0'' = - 2 \gamma_0/5$ for motion perpendicular to
the polarization axis.

In the Doppler or unresolved sideband regime, $\nu \lesssim \gamma_0$, the red and blue sidebands, as well as the carrier transition, overlap, limiting the cooling performance to
\begin{equation}\label{eq:n_ind_dc}
    \bar{n}^{(\text{dc})}_\text{ind}  \approx \frac{\gamma_0 + |\gamma_0 ''|}{4 \nu} \approx \frac{ 7 \gamma_0} {20\nu} > 1,
\end{equation}
which is reached at a slower rate $\Gamma_\text{ind}^{(dc)} \approx 8 \eta^2 \Omega^2 \nu / \gamma_0^2$ to leading order in the small parameter $\nu/\gamma_0$.

Importantly, smaller decay rates yield lower steady-state phonon numbers for all values of $\nu/\gamma_0$, albeit at the cost of reduced cooling rates. This motivates exploring collective subradiant states as a resource for enhanced cooling.

\section{Collective sideband cooling of two atoms}
\label{Section: cooling_2_atoms}

We study the problem of cooling two dipole interacting atoms to their motional ground state. 
This example contains the full conceptual complexity of the problem, while still allowing for analytical solutions in certain regimes. It thus serves as an ideal scenario for illustrating the key concept of our work.
We consider two atoms trapped at a distance $d$ along the $x$-axis, with dipole moments oriented along the $y$-axis. In the following we restrict the motion along the $z$-direction, as illustrated in Fig.~\ref{fig:sketch}(a). Additional configurations and three-dimensional motion are discussed in Appendix~\ref{app:Generalizations}. 

\begin{figure*}[t!]
    \centering
    \includegraphics[width = 2\columnwidth]{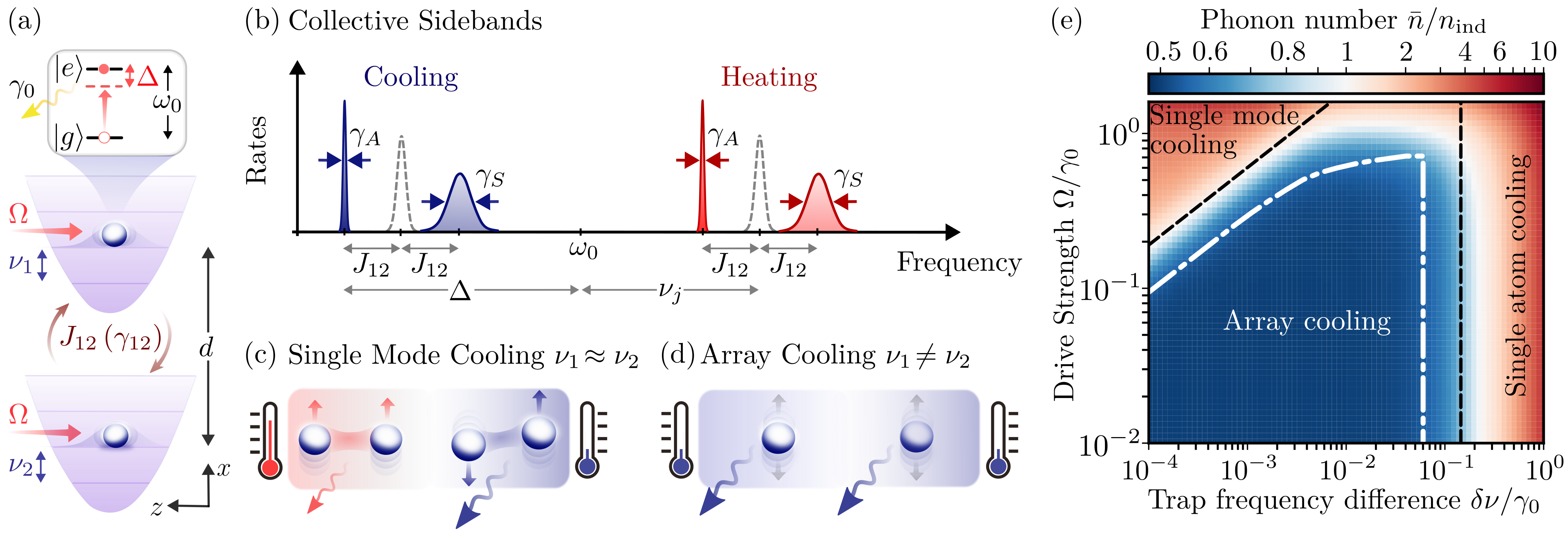}
    \caption{(a) Two harmonically trapped atoms separated by $d<\lambda_0$ are coupled by coherent (dissipative) $J_{12}$ ($\gamma_{12}$) dipole interactions and are driven by an external laser along the direction of motion ($z$).
    (b) Motional sidebands: a cooling and a heating sideband is associated to the symmetric (superradiant) and to the antisymmetric (subradiant) transition. The dashed gray lines indicate the sidebands of noninteracting atoms. (c) For $\nu_1 \approx \nu_2 \approx \nu$, the center-of-mass (relative) motion of both atoms is independently cooled via the symmetric (antisymmetric) cooling sideband to phonon numbers higher (lower) than that of noninteracting atoms. (d) For larger trap frequency difference, both atoms are cooled to lower temperatures via the cooling antisymmetric sideband. (e) Normalized average steady state phonon number, $\bar{n}/\bar{n}_\textit{ind}$, as a function of frequency difference $\delta \nu$ and drive strength $\Omega$ for $d=0.2\lambda_0$, $\nu=20\gamma_0$ and $\Delta \approx -\nu - J_{12}$, obtained by solving the Lamb-Dicke master equation \eqref{eq:ME_Lamb-Dicke}. The dashed black lines correspond to the two conditions in Eq.~\eqref{eq:array_cooling_condition}, which separate the three cooling regimes. The dash-dotted white line corresponds parameters used to compute the critical cooling rate in Fig.~\ref{fig: 2atoms_figure}(b) for $d=0.2\lambda_0$. }
    \label{fig:sketch}
\end{figure*}

The linear response of two dipole-coupled atoms is characterized by two distinctive resonances associated with the excitation of the ground state $\ket{gg}$ to either the symmetric state $\ket{S}=(\ket{ge}+\ket{eg})/\sqrt{2}$ or the antisymmetric state $\ket{A}=(\ket{ge}-\ket{eg})/\sqrt{2}$~\cite{Lehmberg1970b,Beige1999}.
The transition to the symmetric (antisymmetric) states are characterized by a resonance frequency $J_S\equiv\w_0+J_{12} +\eta^2J''_{12}$ ($J_A\equiv\w_0-J_{12}-\eta^2J''_{12}$) and by a collective linewidth
$\gamma_S = \gamma_0 + \gamma_{12} + \eta^2\gamma''_{12}$ 
($\gamma_A = \gamma_0 - \gamma_{12} - \eta^2\gamma''_{12}$).
At short distances ($d < \lambda_0/2$), the symmetric state becomes superradiant with an enhanced decay rate $\gamma_S > \gamma_0$, while the antisymmetric state becomes subradiant with a narrowed linewidth $\gamma_A < \gamma_0$.
In the limit $d \ll \lambda_0$, the dipole-dipole interaction can be expanded in powers of the small parameter $k_0 d$. The coherent coupling scales as $J_{12} \propto (k_0 d)^{-3}$, while the dissipative terms $\gamma_{12}$ and $\gamma_{12}''$ remain finite.
In this limit, the symmetric state decays at $\gamma_S = 2\gamma_0 + \mathcal{O}(k_0^2 d^2)$, whereas the antisymmetric state becomes strongly subradiant, with decay rate 
\begin{equation}
	\gamma_A \approx \frac{\gamma_0}{5} (k_0^2 d^2 + 2 \eta^2) + \mathcal{O}(k_0^4 d^4) \ll \gamma_0,
\end{equation}
and it is ultimately limited by the atoms' zero-point motion in the trap (Sec.~\ref{subsection: Formalism_Effective_ME}).

We consider the atoms driven by a classical plane-wave field propagating along $\mathbf{z}$, such that both atoms experience identical Rabi frequencies: $\Omega_j = \Omega$ and $\Omega_j' = i \Omega$ for $j \in \{1,2\}$. In the following, we assume $\Omega\in \mathbb{R}$ without loss of generality. To zeroth order in $\eta$, this symmetric drive couples exclusively to the symmetric spin state due to conservation of in-plane momentum, $\hat{H}_L^{(0)} = \sqrt{2} \hbar \left( \Omega |S\rangle \langle gg| + \Omega |gg\rangle \langle S| \right)$.
As a result, the steady-state coherences of both atoms depend exclusively on the  properties of the symmetric state, $\avg{\hat{\sigma}_j}_\text{ss} = \Omega / (- \Delta + J_S + i \gamma_S/2)$.

Spin-motion coupling arises from the first-order Lamb-Dicke correction to the driving Hamiltonian,
\begin{equation}\label{eq:HL1_2atoms}
\hat{H}_L^{(1)} \! =\! \frac{i \hbar \Omega}{\sqrt{2}} \! \left[ |S\rangle\langle gg| \left( \! \hat{R}_1 + \hat{R}_2 \! \right) + |A\rangle\langle gg| \left(\! \hat{R}_1 - \hat{R}_2 \! \right) \right] \! + \hc
\end{equation}
Equation~(\ref{eq:HL1_2atoms}) drives sideband transitions at a rate $\eta\W$ to both the symmetric and the antisymmetric spin states, even though the latter is not directly excited by the drive in \eqnref{eq:H_0}. This occurs because in-plane momentum conservation must be satisfied across three subsystems: light, spin, and phonons. As a result, even under a symmetric drive, the antisymmetric (subradiant) spin state becomes relevant when the recoil involves an antisymmetric superposition of motional operators.
Accordingly, both the symmetric and antisymmetric sidebands contribute to the heating and cooling transition rates of each atom,
\begin{eqnarray}
	\mathcal{R}_{jj}^{\pm} &=& \frac{ \mathcal{R}_{Aj}^{\pm}+\mathcal{R}_{Sj}^{\pm}}{2}.
\end{eqnarray}
Here, $\mathcal{R}_{\lambda j}^\pm$ denotes the cooling and heating transition rates associated with each collective spin state $\lambda \in \{S, A\}$, and takes the form
\begin{align}
\label{eq:cool_heat_rates_2_atoms_lambda}
    \mathcal{R}_{\lambda j}^{\pm} & \equiv \frac{\eta_j^2 \Omega^2 \gamma_\lambda}{( - \Delta + J_{\lambda}  \pm \nu_j )^2 + \left( \gamma_\lambda/2 \right)^2 } \nonumber \\
    &+  \frac{ \eta_j^2 \Omega^2 |\gamma_\lambda''|}{(-\Delta + J_{S})^2 + (\gamma_S/2)^2}.
\end{align}
Equation~(\ref{eq:cool_heat_rates_2_atoms_lambda}) has the same structure as the transition rate for independent atoms in Eq.~(\ref{eq:cool_heat_rates_independent}), but with the single-atom frequency and linewidth replaced by those of the collective states.
This means that each collective transition contributes to the cooling (heating) transition rate with a red (blue) sideband as described by the first line of Eq.~(\ref{eq:cool_heat_rates_2_atoms_lambda}) and illustrated in Fig.~\ref{fig:sketch}(b).
The second term accounts for recoil heating due to spontaneous emission from the collective spin states, at rates proportional to $\gamma_\lambda''$. 
Since only the symmetric state is excited by the carrier transition, the recoil-induced Lorentzian appears centered at $\Delta = J_S$ with a width $\gamma_S$, both in $\mathcal{R}_{S j}^{\pm}$ and $\mathcal{R}_{A j}^{\pm}$. Note that for $d\ll\lambda_0$, brighter collective states lead to larger diffusion because $|\gamma_S''| > |\gamma_A''|$.

For independent atoms, the steady state phonon number depends on the linewidth of the dipole transition, with narrower transition yielding lower phonon occupation. 
This indicates that, for two dipole coupled atoms, the narrow antisymmetric ($\gamma_A<\gamma_0$) transition could be exploited to cool atoms to lower motional temperatures. This suggests using a laser detuning
\begin{equation}
    \label{eq: optimal_detuning_2atoms}
    \Delta = -\sqrt{\nu_1^2 + (\gamma_A/2)^2} - J_{12},
\end{equation}
as shown in Fig.~\ref{fig:sketch}(b). 
However, due to motional crosstalks between different atoms both in the form of coherent $V_{ij}$ and dissipative $\mathcal{R}_{ij}^\pm$ mechanical interactions, the minimal achievable phonon numbers $\bar{n}_j$ are not straightforward to predict.
We compute the average steady state population $\bar{n} = \sum_j \bar{n}_j/N$ by solving Eq.~\eqref{eq:ME_Lamb-Dicke} numerically for the case of two atoms and plot the results in Fig.~\ref{fig:sketch}(e).
 We identify three distinct regimes as a function of the Rabi frequency $\Omega$ and trap frequency difference $\delta \nu \equiv |\nu_1 - \nu_2|$.

\begin{figure*}
    \centering   \includegraphics[width = 2\columnwidth]{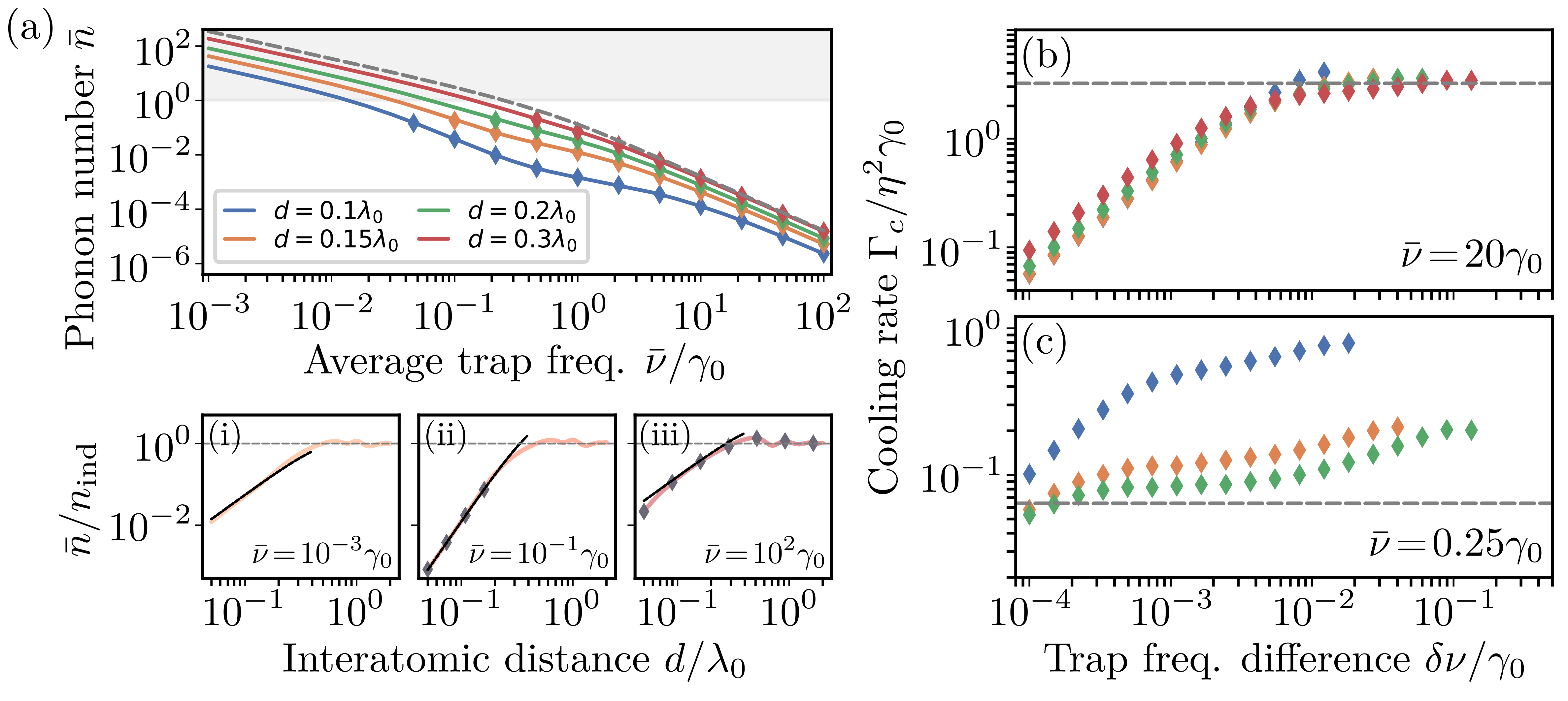}
    \caption{Sideband cooling of two interacting atoms in the array cooling regime for $\eta = 0.02$. (a) Average phonon number $\bar{n}$ as a function of average trap frequency $\bar{\nu}$ for various atomic separations $d$, with $\delta \nu = 10^{-3} \gamma_0$. Solid line are obtained from \eqnref{eq:ME_Motion} while diamond markers are obtained by simulating \eqnref{eq:ME_Lamb-Dicke}. (i-iii) Scaling of $\bar{n}$ with $d$ for various $\bar{\nu}$. The black dashed lines correspond to the analytical estimates given by Eqs.~(\ref{eq: pop_resres})-(\ref{eq: pop_unres_opt}). (b-c) Critical cooling rate $\Gamma_c$ as a function of $\delta \nu$ for various $d$ and for (b) $\bar{\nu}=20\gamma_0$ and (c) $\bar{\nu}=0.25 \gamma_0$. For the case $d=0.2\lambda_0$ shown in panel (b), the driving strength corresponding to the critical cooling rate is indicated by the white dash-dotted line in \figref{fig:sketch}(e). In all panels, the dashed gray lines correspond to the results for noninteracting atoms, while the solid lines and diamond markers are respectively obtained by solving the effective master equation~\eqref{eq:ME_Motion} and the full model including the spin dynamics. For the numerical simulation in panel (a) we set $\Omega/\gamma_0=10^{-3}$.}
    \label{fig: 2atoms_figure}
\end{figure*}

\subsection{Single-mode cooling}

The first regime emerges for small trap frequency differences,
\begin{equation}
\delta \nu \ll  \frac{2 \eta_j^2 \Omega^2}{\gamma_A}\quad \, \forall j.\label{eq:FreqDiff_Coupling}
\end{equation}
Combined with the weak spin-motion coupling condition $\eta \Omega \ll \gamma_A$, Eq.~(\ref{eq:FreqDiff_Coupling}) also implies $\delta \nu \ll \gamma_A/2$. The trap frequency difference is much smaller than the linewidths of the sidebands, hence $\nu_j \approx \bar{\nu}$, $\eta_j\approx \eta$, and the collective rates in Eq.~(\ref{eq:cool_heat_rates_2_atoms_lambda}) are approximately equal for both atoms, $\mathcal{R}_{\lambda 2}^{\pm} \approx \mathcal{R}_{\lambda 1}^{\pm} \equiv \mathcal{R}_{\lambda}^{\pm}$ for $\lambda \in \{ S, A \}$.

At the optimal detuning, \eqref{eq: optimal_detuning_2atoms}, the dissipative motional coupling rate in Eq.~(\ref{eq:ME_Motion}) becomes
\begin{equation}
\label{eq:R_12}
    \mathcal{R}_{12}^{-} \approx \frac{\mathcal{R}_S^{-} - \mathcal{R}_A^{-}}{2} \approx -2 \frac{\eta^2 \Omega^2}{\gamma_A}.
\end{equation}
In the last passage we neglected the contribution from the symmetric spin state which is suppressed due to its larger decay rate $\gamma_S$ and its detuning from the drive caused by dipole-induced level splitting $2 J_{12}$. As a result, $\mathcal{R}_{12}^{-}$ is dominated by the antisymmetric (subradiant) cooling sideband.

At the optimal detuning, the dissipative motional coupling associated to cooling, $\mathcal{R}_{12}^-$, is larger than that associated to heating, $\mathcal{R}_{12}^+$, and over the coherent motional coupling $V_{12}$~\footnote{For example, this can be easily in the regime $\bar{\nu} \gg \gamma_0$, where $R_{12}^+ \propto \eta ^2 \Omega^2 \gamma_0/\bar{\nu}^2 $ and $V_{12} \propto \eta^2 \Omega^2 / \bar{\nu}$.}. Thus, Eq.~(\ref{eq:FreqDiff_Coupling}) can be interpreted as the condition under which the motional degrees of freedom of both atoms are resonantly coupled. Then, the effective master equation for the atomic motion becomes diagonal when expressed in terms of the center of mass and relative motional modes, defined by the operators \( \hat{b}_S = (\hat{b}_1 + \hat{b}_2)/\sqrt{2} \) and \( \hat{b}_A = (\hat{b}_1 - \hat{b}_2)/\sqrt{2} \), respectively. The problem thus reduces to two independent instances of sideband cooling, where the center-of-mass and relative motion are respectively cooled by the symmetric and antisymmetric spin states at rates $\mathcal{R}_S^\pm$ and $\mathcal{R}_A^\pm$, as illustrated in Fig.~\ref{fig:sketch}(c).

In the resolved sideband regime ($\bar{\nu} \gg \gamma_0$), the steady-state phonon numbers of the collective motional modes, $\bar{n}_\lambda = \mathcal{R}_\lambda^{+} / (\mathcal{R}_\lambda^{-} - \mathcal{R}_\lambda^{+})$, are given by
\begin{align}
\label{eq: pops_2atoms_equaltrap}
    \bar{n}_A &\approx \frac{\gamma_A^2}{16 \bar{\nu}^2} \left( 1 + 4 \frac{|\gamma_A^{''}|}{\gamma_A} \right), \\
    \bar{n}_S &\approx \frac{\gamma_S^2 + 16 J_{12}^2}{16 \bar{\nu}^2} \left( 1 + 4 \frac{|\gamma_S^{''}|}{\gamma_S} \right).
\end{align}
Compared with the steady-state phonon number for noninteracting atoms in the resolved sideband regime, given by \eqnref{eq:n_ind_sc}, the occupations $\bar{n}_S$ and $\bar{n}_A$ are modified by the decay properties of the collective spin states. At short distances $d < \lambda_0/2$, $\gamma_A<\gamma_0$ and hence $\bar{n}_A < \bar{n}_{\text{ind}}^{\text{(sc)}}$. In contrast, the symmetric state is superradiant, leading to $\bar{n}_S > \bar{n}_{\text{ind}}^{(\text{sc})}$.

While collective interactions enable enhanced cooling of one collective motional mode (in this case, the relative motion), the average phonon number of the system is necessarily larger than that of two noninteracting atoms,
\begin{equation}
    \bar{n} = \frac{\bar{n}_A + \bar{n}_S}{2} \geq \bar{n}_{\text{ind}}^{(\text{sc})},
\end{equation}
indicating no net improvement in the overall cooling performance~\footnote{The inequality above can be verified analytically in the absence of recoil heating. In that limit, it is enforced by the relation $(\gamma_A^2 + \gamma_S^2)/2 = \gamma_0^2 + \gamma_{12}^2 \geq \gamma_0^2$,
which follows from the triangle inequality, and by the frequency splitting \( 2J_{12} \) between the cooling sidebands of the symmetric and antisymmetric spin states.}. This is why we refer to this regime as \emph{single-mode cooling}: only one of the collective motional modes benefits from the subradiant enhancement. It corresponds to the upper-left corner of Fig.~\ref{fig:sketch}(e).

\subsection{Array cooling}

The second regime is characterized by the following condition
\begin{equation}\label{eq:array_cooling_condition}
	\frac{2 \eta_j^2 \Omega^2}{\gamma_A} \ll \delta \nu \ll \frac{\gamma_A}{2}.
\end{equation}
The left-half of the inequality in \eqnref{eq:array_cooling_condition} ensures that atoms are effectively decoupled as $\mathcal{R}_{ij}$ and $V_{ij}$ in \eqnref{eq:ME_Motion} can be neglected under the rotating wave approximation when $i\neq j$.
The effective master equation thus becomes diagonal in the atomic basis, and the steady-state phonon numbers reduce to
\begin{equation}
\label{eq: SI_av_pop_arraycooling_general}
    \bar{n}_j = \frac{\mathcal{R}_{jj}^{+}}{\mathcal{R}_{jj}^{-} - \mathcal{R}_{jj}^{+}} = \frac{\mathcal{R}_{A j}^{+} + \mathcal{R}_{S j}^{+}}{\mathcal{R}_{A j}^{-} + \mathcal{R}_{S j}^{-} - \mathcal{R}_{A j}^{+} - \mathcal{R}_{S j}^{+}}.
\end{equation}
The right-half of \eqnref{eq:array_cooling_condition} ensures that the antisymmetric cooling sideband condition can be satisfied for both atoms and thus $\mathcal{R}_{\lambda j}^{\pm} \equiv \mathcal{R}_{\lambda}^{\pm}$. In this case, both atoms reach the same steady-state phonon number, $\bar{n} \approx \bar{n}_1 \approx \bar{n}_2$, which depends only on the average trap frequency $\bar{\nu}$.

Crucially, the cooling dynamics of each individual atom are determined by the properties of the collective spin states, particularly of the subradiant state, such that $\mathcal{R}_{jj}^- \approx \mathcal{R}_{A}^-/2$ [\figref{fig:sketch}.(d)]. This leads to a lower average phonon number compared with the noninteracting case over a broad range of average trap frequencies $\bar{\nu}$, as shown in Fig.~\ref{fig: 2atoms_figure}(a). 
To quantify the enhancement in cooling performance, we compute $\bar{n} / \bar{n}_{\text{ind}}$ and identify three distinct scenarios depending on the magnitude of $\bar{\nu}$.

(i) For $\bar{\nu}\gg \gamma_S \approx \gamma_0$, the sidebands are resolved both in the case of interacting and non-interacting atoms. The cooling transition rate is then dominated by the subradiant cooling sideband, yielding $\mathcal{R}_{jj}^{-}  \approx 2 \eta^2 \Omega^2 /\gamma_A$. To compute the heating transition rate $\mathcal{R}_{jj}^+$, we need to distinguish two limiting cases based on the strength of the light-induced energy shift $J_{12}$.

If $\bar{\nu} \gg |J_{12}|$, both spin states contribute comparably to the heating rate. Using the relations $\gamma_A + \gamma_S = \gamma_0$ and $\gamma_A '' + \gamma_S'' = \gamma_0''$, it can be approximated as $\mathcal{R}_{jj}^{+}  \approx 13\eta^2 \Omega^2 \gamma_0 / 20 \nu^2$. The resulting steady-state phonon number $\bar{n} \approx 13\gamma_A \gamma_0/40\bar{\nu}^2$ normalized to the case of independent atoms reads
\begin{equation}
\label{eq: pop_resres}
     \frac{\bar{n}}{\bar{n}_\text{ind}^{(\text{sc})}} \approx 2\frac{\gamma_A}{\gamma_0}.
\end{equation}

If $|J_{12}| \gg \bar{\nu}$, the heating sideband of the antisymmetric spin state is detuned from the drive by $2 \bar{\nu}$, while the heating sideband of the symmetric spin state and the Lorentzian associated to recoil are detuned by $ 2 \bar{\nu} + 2 J_{12} \gg 2 \bar{\nu}$ and $\bar{\nu} + 2 J_{12} \gg 2\bar{\nu}$, respectively. As a result, only the heating sideband of the antisymmetric spin state significantly contributes to heating, $\mathcal{R}_{jj}^{+}  \approx \eta^2 \Omega^2 \gamma_A / 8 \nu^2$. The steady-state phonon number $\bar{n} \approx \gamma_A^2/16\bar{\nu}^2$ normalized to the case of independent atoms reads
\begin{equation}
\label{eq: pop_unres_res}
     \frac{\bar{n}}{\bar{n}_\text{ind}^{(\text{sc})}} \approx \frac{5}{13}\frac{\gamma_A^2}{\gamma_0^2}.
\end{equation}
This regime typically requires very small separations between the atoms, such that $J_{12} \gg \gamma_0$. The transition from Eqs.~(\ref{eq: pop_resres}) and~(\ref{eq: pop_unres_res}) is showcased in the deviation of the numerically obtained phonon number from Eq.~(\ref{eq: pop_resres}) in Fig.~\ref{fig: 2atoms_figure}(a,iii). 

(ii) For $\gamma_A\ll\bar{\nu}\lesssim \gamma_0$, only the sidebands associated to the antisymmetric subradiant spin rate are resolved; that is, ground-state cooling is only possible for dipole-interacting atoms. The denominator in Eq.~(\ref{eq: SI_av_pop_arraycooling_general}) remains dominated by the antisymmetric cooling rate, $\mathcal{R}_A^{-}$. For $d \geq 0.1\lambda_0$, we typically have $|J_{12}| \geq \gamma_0 \geq \bar{\nu}$, such that the heating rate is also mainly given by the properties of the antisymmetric spin state. The resulting steady-state phonon number $\bar{n} \approx \gamma_A^2/16\bar{\nu}^2$ normalized to the case of independent atoms in the the Doppler-cooling regime reads
\begin{equation}
\label{eq: pop_res}
\frac{\bar{n}}{\bar{n}^{(\text{dc})}_\text{ind}} \approx \frac{5}{28}\frac{\gamma_A^2}{\bar{\nu}\gamma_0} .
\end{equation}
Remarkably, ground-state cooling of the two interacting atoms is achievable in this scenario, even if it is not possible for a single independent atom.

(iii) For $\bar{\nu} \ll \gamma_A/2$,
neither the collective nor the independent sidebands are resolved. Nonetheless, the cooling and heating rates are still dominated by the antisymmetric spin state, and can be approximated as $\mathcal{R}_{jj}^{\pm} \approx  \eta^2 \Omega^2 / \gamma_A \pm  2 \eta^2 \Omega^2 \bar{\nu} / \gamma_A^2$.
Then, the steady-state phonon number $\bar{n} \approx \gamma_A/4\bar{\nu}$ normalized  to the case of independent atoms in the the Doppler-cooling regime reads
\begin{equation}
\label{eq: pop_unres_opt}
    \frac{\bar{n}}{\bar{n}^{(\text{dc})}_\text{ind}} \approx \frac{5}{7}\frac{\gamma_A}{\gamma_0} .
\end{equation}
While the phonon number remains lower than for noninteracting atoms, it exceeds unity—implying that ground state-cooling is not achievable.

In the regime defined by \eqnref{eq:array_cooling_condition}, the steady-state phonon number scales with the subradiant decay rate $\gamma_A$ across all values of $\bar{\nu}$, enabling cooling of all atoms to temperatures lower than those achievable with non-interacting atoms. We thus refer to it as the \emph{array cooling} regime [see central blue region of Fig.~\ref{fig:sketch}(e)]. As shown in Fig.~\ref{fig: 2atoms_figure}(a,i)-\ref{fig: 2atoms_figure}(a,iii), the predictions of Eqs.~(\ref{eq: pop_resres})--(\ref{eq: pop_unres_opt}) are in good agreement with numerical results obtained from the effective master equation~(\ref{eq:ME_Motion}) (solid lines), as well as from the full model including the spin dynamics (diamond markers) across a range spacings $d$ and trap frequencies $\bar{\nu}$.

We note that in the configuration considered here, where $J_{12} > 0$, the antisymmetric cooling sideband never overlaps with any heating sidebands [see Fig.~\ref{fig:sketch}(b)]. As a result, enhanced cooling is achieved across the entire range of trap frequencies $\bar{\nu}$, as shown in Fig.~\ref{fig: 2atoms_figure}(a). In contrast, for $J_{12} < 0$ (e.g., for dipole polarization aligned along the atomic chain), the antisymmetric cooling sideband is shifted to higher frequencies and can become resonant with heating sidebands at specific values of interatomic spacing $d$ and trap frequency $\bar{\nu}$. This coalescence of sidebands leads to an increase in the steady-state phonon number and can even prevent cooling as discussed in Appendix~\ref{app:Generalizations}. 

Let us now examine the impact of dipole-dipole interactions on the cooling rate $\Gamma$. In the array cooling regime, when the subradiant sidebands are resolved, the cooling rate is approximately given by the cooling transition rate associated to the antisymmetric spin state
\begin{equation}
\Gamma \approx \frac{(\eta \Omega)^2}{\gamma_A}.
\end{equation}
Stronger driving fields $\Omega$ result in faster cooling. However, for $\eta$ fixed, $\Omega$ is constrained by either Eq.~(\ref{eq:array_cooling_condition}), --which ensures that the system to remain within the array cooling regime, --or the weak spin-motion coupling condition $\eta \Omega \ll \gamma_A$. Violating either condition leads to higher steady-state phonon numbers.

To quantify these limits, we numerically solve the full coupled spin-motion dynamics equation~\eqref{eq:ME_Lamb-Dicke} and extract the cooling rate $\Gamma$ by fitting the time evolution of the average phonon number $\bar{n}$. We define the critical cooling rate $\Gamma_c$ as the value of $\Gamma$ obtained at the maximum drive strength $\Omega$ that results in a steady-state phonon number at most 10\% larger than the optimal prediction from Eq.~(\ref{eq:ME_Motion}) [see white line in Fig.~\ref{fig:sketch}(e)]. These results are shown in Fig.~\ref{fig: 2atoms_figure}(b) for $\bar{\nu} = 20 \gamma_0$. 
For small trap frequency differences $\delta \nu \ll \gamma_A / 2$, the maximum $\W$ is limited by the array cooling condition in \eqnref{eq:array_cooling_condition}, hence $\Gamma_c\propto \delta\nu$. In this case, the cooling rate is smaller than for non-interacting atoms.
As $\delta \nu$ increases, the dominant constraint on $\Omega$ becomes the weak-coupling condition, and $\Gamma_c$ saturates to values comparable to those of non-interacting atoms. If the frequency difference becomes too large, $\delta \nu \sim \gamma_A / 2$, the system enters the single-atom cooling regime. In this case, it is no longer possible to find a drive strength that yields an average phonon number close to the optimal prediction, resulting in the absence of markers in Fig.~\ref{fig: 2atoms_figure}(b) at large $\delta \nu$. 
Figure~\ref{fig: 2atoms_figure}(c) shows the results for $\bar{\nu} = 0.25 \gamma_0$, where motional sidebands are only resolved via the collective subradiant mode [see regime (ii) above]. In this scenario, the subradiant collective spin state enables significantly lower steady-state phonon numbers and faster cooling rates than in the noninteracting case, demonstrating a clear advantage of collective interactions.
Additional detail on the numerical optimization procedure used to obtain the critical cooling rate are given in \appref{App: Optimization_Cooling_Rates}.

\subsection{Single-atom cooling}

The third and final regime in Fig.~\ref{fig:sketch}(e) corresponds to the condition
\begin{equation}
\label{eq:App_condition_singleatom}
    \delta \nu \gg \frac{\gamma_A}{2}.
\end{equation}
Here, the trap frequency difference exceeds the width of the sidebands, so the cooling and heating rates $\mathcal{R}_{\lambda j}^{\pm}$ differ between atoms. A single driving field cannot simultaneously address the antisymmetric cooling sideband of both atoms. 
If the drive is tuned to the antisymmetric cooling sideband of atom one, $\Delta = - \sqrt{\nu_1^2 + (\gamma_A/2)^2} - J_{12}$, then atom one is cooled to the same phonon number as in the array cooling regime. However, cooling of the second atom becomes inefficient, and its phonon number typically exceeds that of non-interacting atoms~\footnote{For example, in the resolved sideband regime $\nu_{1,2} \gg \gamma_0$, assuming $\nu_{1,2} \gg |J_{12}|$ and $\delta \nu \ll \nu_2$, the cooling rate for the second atom is suppressed to $\mathcal{R}_{A2}^{-} + \mathcal{R}_{S2}^{-} \approx \frac{\eta^2 \Omega^2 \gamma_A}{\gamma_A^2 + 4 \delta \nu^2}$. Then, its steady-state phonon number increases to $\bar{n}_2 \approx \bar{n}_1 (1 + 4 \delta \nu^2 / \gamma_A^2)$.}. This \emph{single-atom cooling} regime corresponds to the right edge of Fig.~\ref{fig:sketch}(e).

This regime is of importance if one aims at cooling a single atom using the surrounding atoms as coolant to reach lower final temperature for the target atoms. We consider such a situation below for the case of an atom surrounded by a ring of emitters.

\section{Collective sideband cooling of atomic arrays}
\label{Section: cooling_N_atoms}

\begin{figure}
    \centering    \includegraphics[width=\columnwidth]{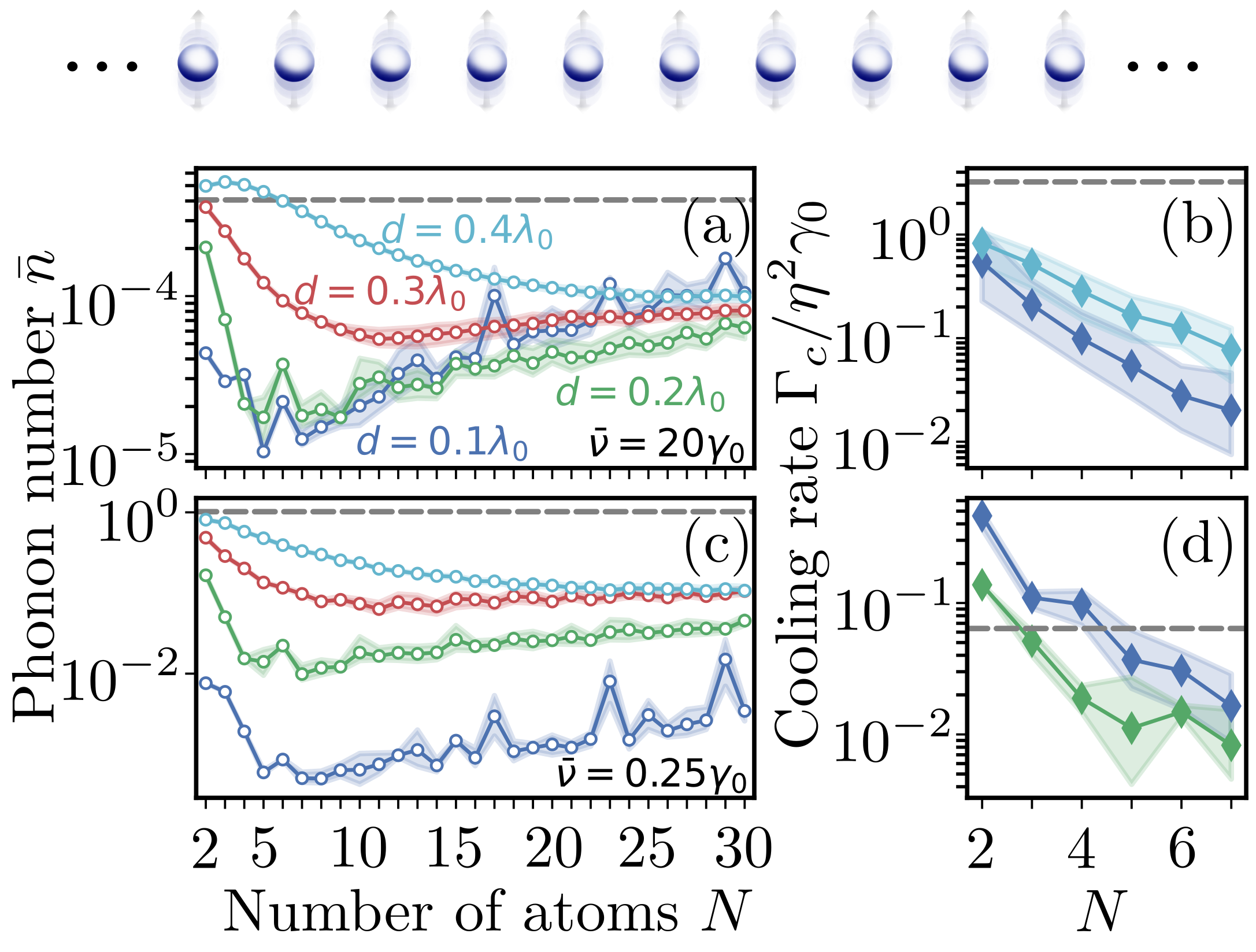}
    \caption{Sideband cooling of a chain of $N$ interacting atoms.
    (a) Steady state phonon number per atom $\bar{n}$ and (b) critical cooling rate $\Gamma_c$ as a function of $N$ for various spacings $d$. The trap frequencies are normally distributed with $\bar{\nu}=20 \gamma_0$ and  $\sigma  = 10^{-3} \gamma_0$. The drive is resonant with the subradiant spin mode resulting in the smallest $\bar{n}$. The cooling rates for $d=0.2\lambda_0$ and $d=0.3\lambda_0$ exhibit similar trends and values. (c-d) Analogous for $\bar{\nu}=0.25 \gamma_0$. In all panels, we plot the median (markers), as well as the $25$-th to $75$-th percentile region (shaded area) obtained by averaging over $300$ realizations. Diamond markers and empty circles correspond respectively to the solution of \eqnref{eq:ME_Lamb-Dicke} and \eqref{eq:ME_Motion}. The gray dashed lines correspond to the case of non-interacting atoms $d \gg \lambda_0$. The solid lines are included as guides to the eye. For small systems of up to seven atoms, the steady-state phonon numbers obtained by solving the full system agree well with those predicted by the effective model in Eq.~(\ref{eq:ME_Motion}) (data not shown).}
    \label{fig: N_atoms_gradient}
\end{figure}

The fundamental concepts developed for two interacting atoms can be extended to larger systems of $N$ atoms. In this case, all $N$ collective spin states of the system contribute to the rates $\mathcal{R}_{ij}^{\pm}$ in Eq.~(\ref{eq:ME_Motion}). Assuming the collective spin modes are delocalized and overlap with each atom with equal amplitude $\sim 1/\sqrt{N}$—as is the case in rings or extended atomic arrays—the cooling and heating rates for atom $j$ can be approximated as
\begin{equation}\label{eq:Rates_N_atoms}
    \mathcal{R}_{jj}^\pm \approx \frac{1}{N} \sum_\lambda \mathcal{R}_{\lambda j}^\pm,
\end{equation}
where $\mathcal{R}_{\lambda j}^\pm$ are the rates associated with each collective spin state $\lambda$, given in Eq.~(\ref{eq:cool_heat_rates_2_atoms_lambda}). As in the two-atom case, dipole-dipole interactions lead to different energy shifts for each collective spin states and its associated sidebands. A single monochromatic laser field can only address one (or at most a few) of these sidebands resonantly. Consequently, only those spin states whose sidebands are close to resonance will significantly contribute to \eqnref{eq:Rates_N_atoms}.

\subsection{Cooling an $N$-atom array}

Enhanced cooling of all atoms—\ie a net improvement in the overall cooling performance—requires the driving field to be resonant with the cooling sideband of a subradiant state that is delocalized across the entire array. This is because only atoms with non-zero amplitude in the targeted collective spin state can be effectively cooled. Moreover, the system must operate in the \emph{array cooling} regime. This requires that the trap frequency differences satisfy
\begin{equation}
\label{eq: con_array_cooling_Natoms}
|\nu_i - \nu_j| \gg \frac{2 \eta_j^2 \Omega^2}{\gamma_\lambda} \quad \,\forall i,j,\lambda.
\end{equation}
Importantly, achieving this regime does not necessarily demand precise control over individual trap frequencies; it can instead arise naturally from the inherent fluctuations and disorder present in experimental implementations.

To illustrate this point, we assume the trap frequencies $\nu_j$ are normally distributed with mean $\bar{\nu}$ and variance $\sigma^2$, and we address the cooling sideband that yields the lowest steady-state phonon number.~\footnote{For a subradiant spin state with collective linewidth $\gamma_\lambda$ and energy shift $J_\lambda$, the optimal detuning of the laser drive is $\Delta = -\sqrt{\bar{\nu}^2 + (\gamma_\lambda/2)^2 } + J_\lambda$.} Figures~\ref{fig: N_atoms_gradient}(a) and (c) show enhanced cooling for subwavelength chains with up to $N = 30$ atoms, in both the resolved ($\bar{\nu} = 20\gamma_0$) and unresolved ($\bar{\nu} = 0.25\gamma_0$) regimes (analogous results hold for other geometries, see Appendix~\ref{app:Geometries}). 
As in the two-atom case, the enhancement is more pronounced in the regime where the collective subradiant sidebands are resolved but the individual atomic sidebands are not. Notably, we identify conditions under which the phonon number is suppressed by more than two orders of magnitude. As $N$ increases, the steady-state phonon number begins to rise. This can be attributed to two main effects. First, the size of the array cooling region shrinks, since it becomes increasingly difficult to satisfy the condition in Eq.~(\ref{eq: con_array_cooling_Natoms}) for all frequency differences $|\nu_i - \nu_j|$. Second, while only a few collective spin states significantly contribute to cooling, a larger fraction contributes to heating. 

In Figures.~\ref{fig: N_atoms_gradient}(b) and \ref{fig: N_atoms_gradient}(d), we show the optimal cooling rate $\Gamma_c$, defined analogously to the two-atom case. We observe that $\Gamma_c$ decreases with increasing atom number, regardless of the value of $\bar{\nu}$. We attribute this behavior to the narrowing of the smallest decay rates for a few atoms,~\footnote{For larger number of atoms the subradiant linewidth do not scale with the system size as arguend in \ref{Section: System_Model}, see \eqnref{eq:gamma_sub_lower_bound}.} as well as to the fact that only a few (delocalized subradiant) spin states can significantly contribute to cooling due to the collective energy shifts. 

Finally, it is important to note that, as discussed in the two-atom case, certain polarizations can cause heating and cooling sidebands to become resonant when $\bar{\nu}$ is comparable to the dipole-induced energy splitting between collective spin states (Appendix~\ref{app:Generalizations}). The coalescence between a heating and a cooling sidebands can result in suppression of the cooling and overall heating of the system. To ensure enhanced cooling, it is therefore essential to avoid such coalescence.

\begin{figure*}
    \centering 
    \includegraphics[width = \textwidth]{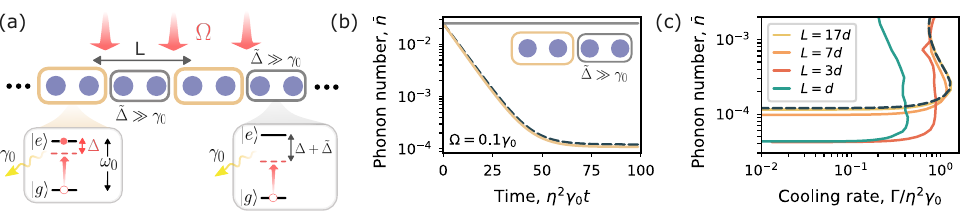}
    \caption{ Protocol for sequential cooling. (a) Distant blocks of a few atoms (orange boxes) are cooled simultaneously, while the spin transition of the remaining atoms (gray boxes) is detuned by and additional shift $\tilde{\Delta} \gg \gamma_0$. Blocks of resonant atoms that are sufficiently far apart ($L \gtrsim \lambda_0$) are cooled independently. (b) Cooling of a four-atom chain with spacing $d=0.15 \lambda_0$, where the last two atoms are far detuned by $\tilde{\Delta} = - 40 \gamma_0$ (\ie $L=d$). Light on resonance with the cooling sideband associated to the first two atoms (orange trace) leads to cooling with the same properties as for a chain of two atoms (black dashed line). The phonon number of the detuned atoms remains constant (gray line). (c) Steady state phonon number $\bar{n}$ versus cooling rate $\Gamma$ for four atoms located along the $x$-axis at positions $x_1=0$, $x_2=d$, $x_3 = L + d$ and $x_4 = L+ 2d$ for $d=0.15 \lambda_0$. The different pairs of $\bar{n}$ and $\Gamma$ are obtained by scanning the drive strength $\Omega$. The cooling properties of a two-atom chain (dashed lines) is recovered at large $L$. We consider $\eta = 0.02$ and $\nu_j = \bar{\nu} + \delta \nu (j - 3)$ with $\bar{\nu} = 20 \gamma_0$ and $\delta \nu = 10^{-3} \gamma_0$. } 
    \label{fig: coolingrates_2}
\end{figure*}

Faster cooling of large atomic arrays can be achieved by sequentially addressing and cooling small groups of atoms. To this end, the array is divided into blocks of $N_b$ atoms, as illustrated in Fig.~\ref{fig: coolingrates_2}(a). A subset of these blocks is designated as target blocks (orange boxes), while the atoms in the remaining blocks (grey boxes) are detuned from the drive by a large amount $|\tilde{\Delta}| \gg \gamma_0$. To ensure fast cooling, the separation between adjacent target blocks must satisfy $L \gtrsim \lambda_0$. This protocol leverages two key features.

First, the detuned atoms have a negligible effect on the radiative properties and cooling dynamics of the target atoms. This is illustrated in Fig.~\ref{fig: coolingrates_2}(b), where two resonant atoms embedded in a four-atom chain (orange trace) are cooled to the same steady-state phonon number and at the same rate as an isolated two-atom array (black dashed trace). Importantly, the dynamics of the detuned atoms are significantly slower than those of the resonant (target) atoms, so that their phonon numbers remain effectively constant throughout the cooling process (gray trace).

Second, when target blocks are sufficiently separated, their radiative properties closely match those of an isolated block. This is illustrated in Fig.~\ref{fig: coolingrates_2}(c), where we consider two blocks of two atoms each, separated by a distance $L$. By scanning the drive strength $\Omega$, we extract the cooling rate $\Gamma$ as a function of the steady-state phonon number $\bar{n}$ for different values of $L$. As $L$ increases, the cooling performance approaches that of an isolated two-atom chain (black dashed trace), and becomes significantly faster than in the case of four adjacent atoms (green trace for $L = d$). Naturally, the phonon number $\bar{n}$ also converges to that of the two-atom system.

Together, these features allow multiple blocks to be cooled simultaneously to the same phonon number and at the same rate as a single block. By repeating the process for the initially detuned blocks, large arrays of $N$ atoms can be cooled sequentially in a number of steps $\sim \lambda_0 / (N_b d)$ that does not scale with the total system size. This significantly reduces the total time required for cooling.

Alternative strategies to accelerate cooling include engineering spatial atomic arrangements that support multiple subradiant spin states with (near-)degenerate resonance frequencies, allowing their cooling sidebands to be simultaneously excited with a single laser. Another approach involves applying multiple driving fields, each tuned to the cooling sideband of a different subradiant state. While this can enhance the overall cooling rate, it also increases heating rates and may ultimately result in higher steady-state phonon numbers. Finally, the cooling process can be sped up by dynamically tuning the interatomic spacing~\cite{Greiner_accordion_dipolar_2023}. Starting with $d \gg \lambda_0$, the atoms behave approximately as independent emitters, enabling rapid initial cooling of large arrays to the steady-state phonon number associated with non-interacting atoms. The spacing can then be reduced to enter the regime of collective cooling and reach lower phonon numbers at the reduced speed. 

\subsection{Single-atom cooling of an emitter in a ring} \label{Section: Ring_Cooling}

\begin{figure}
    \centering 
    \includegraphics[width=\columnwidth]{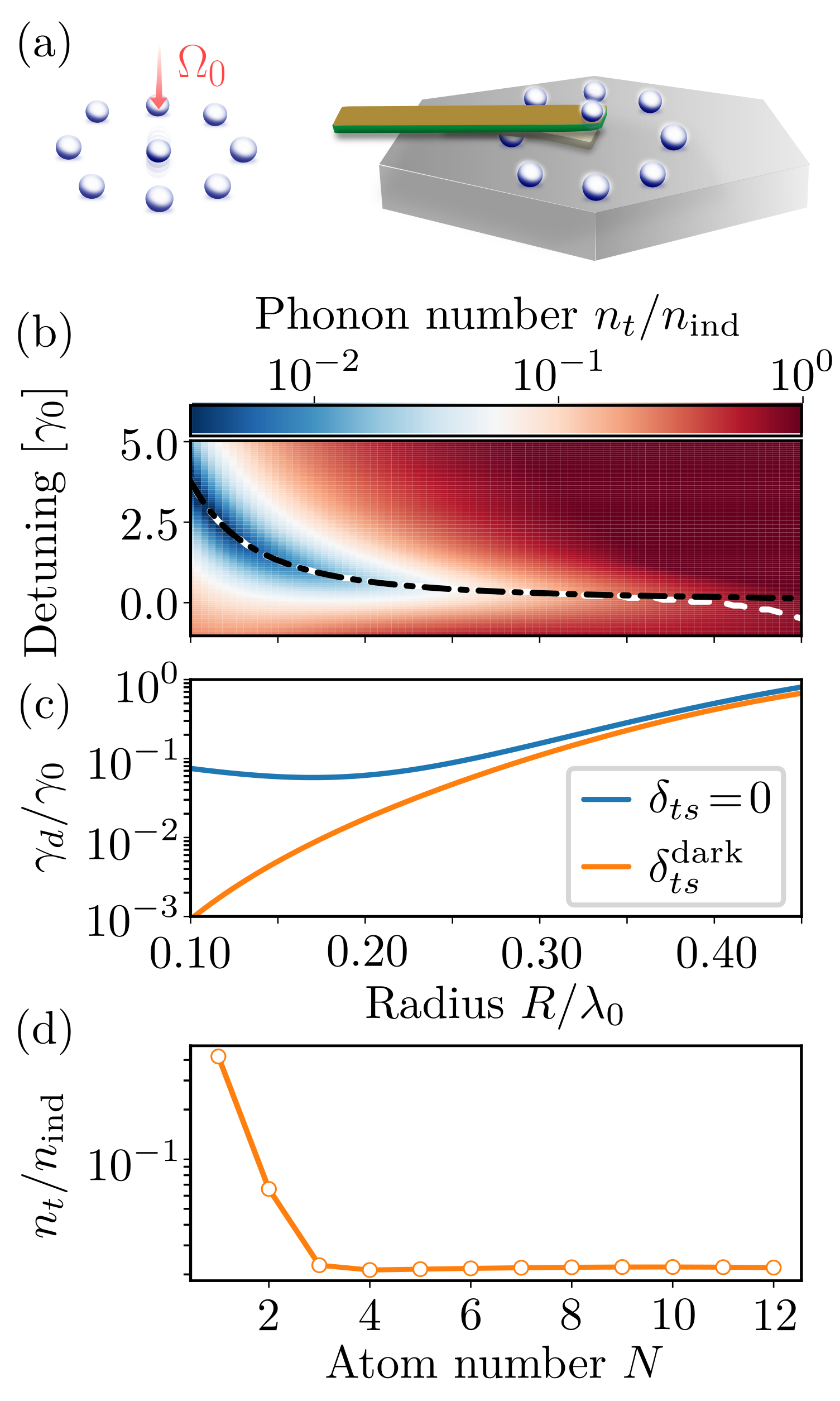}
    \caption{Collective cooling of a single atom. (a) Ground state cooling of a moving quantum emitter (in free space or embedded in a mechanical resonator) via the light-induced dipole-dipole interactions with a nearby ring of $N$ emitters and radius $R$. (b) Steady state phonon number of the target resonator as a function of radius $R$ and detuning $\delta_{tS}$ for a seven-atom ring. 
    The black dash-dotted line and white dashed line indicated respectively \eqnref{eq: optimal_detuning} and the detuning leading to the minimum steady-state population. (c) Decay rate of the subradiant spin mode as a function of radius for $\delta_{tS}=0$ and $\delta_{tS}^\mathrm{dark}$. (d) Steady state phonon number of the target emitter as a function of $N$ for a ring of radius $R=\lambda_0/5$ for $\delta_{tS}=\delta_{tS}^\mathrm{dark}$. In all panels, we consider $\eta_t = 0.02$, $\nu_t = 50 \gamma_c$ and drive perpendicular to the plane defined by the ring with circular in-plane polarization.} 
    \label{fig: RingCooling}
\end{figure}

We now investigate how the \emph{single-atom cooling} regime generalizes beyond the two-atom case.
To this end, we consider a single target emitter embedded in an ensemble of $N$ auxiliary emitters, as illustrated in Fig.~\ref{fig: RingCooling}(a). 
In what follows, we identify the conditions under which auxiliary atoms enhance cooling of the target emitter, and quantify the resulting improvement in cooling performance. 

We assume the target atom and the auxiliary atoms to be trapped at frequencies $\nu_t$ and $\nu_a$ respectively. If $|\nu_t - \nu_a|\gg V_{tn},\mathcal{R}_{tn}^\pm$, where $n=1,\ldots N$ labels the auxiliary atoms, the interaction becomes far off-resonant and can be neglected within the rotating wave approximation. In this regime, the target atom evolves independently,
\begin{equation}
\label{eq: EOM_central_atom}
    \frac{d}{dt} \langle \hat{b}_t^\dagger \hat{b}_t \rangle = - (\mathcal{R}_{tt}^- - \mathcal{R}_{tt}^+) \langle \hat{b}_t^\dagger \hat{b}_t \rangle +  \mathcal{R}_{tt}^+,
\end{equation}
and its steady-state phonon number $n_t$ can be significantly lower than for an isolated atom due to the collective nature of photon emission. It is worth noting that Eq.~(\ref{eq: EOM_central_atom}) remains valid even when the auxiliary emitters are stationary. This extends the applicability of our protocol beyond cold atom arrays to hybrid systems, such as mechanical oscillators coupled to quantum emitters. A notable example is a nitrogen-vacancy (NV) center in diamond embedded in a vibrating cantilever or nanobeam~\cite{cantilever_Arcizet,cantilever_perdriat,cantilever_Maletisnky,cantilever_Rabl,cantilever_Loncar}, which realizes a spin-motion interaction analogous to that considered here.

We consider the symmetric scenario of Fig.~\ref{fig: RingCooling}(a), where the target atom is at the center of a ring of $N$ auxiliary atoms. For motion perpendicular to the ring and an atomic polarization that preserves the symmetry (\eg circular in-plane or perpendicular), the target atom couples only to the symmetric ring spin mode $\ket{S} = \sum_{n=1}^N \hat{\sigma}_n^\dagger \ket{0}/\sqrt{N} \equiv \hat{\sigma}_S^\dagger \ket{0}$, with collective frequency shift $J_S = \sum_{n=2}^N J_{1n}$ and decay rate $\Gamma_S = \sum_{n=1}^N \Gamma_{1n}$. Then, the cooling and heating transition rates $\mathcal{R}_{tt}^\pm$ only depend on the eigenstates of the non-Hermitian spin Hamiltonian
\begin{align}
\label{eq: nonHermitian_Ham_ring}
    \hat{H}_S &= - \Delta \left( \hat{\sigma}_t^\dagger \hat{\sigma}_t + \hat{\sigma}_S^\dagger \hat{\sigma}_S \right) + \left( \delta_{tS} - i \frac{\gamma_0}{2} \right) \hat{\sigma}_t^\dagger \hat{\sigma}_t \nonumber \\
    & + \! \left( \! J_S \! - \! i \frac{\Gamma_S}{2} \! \right) \! \hat{\sigma}_S^\dagger \hat{\sigma}_S + \left( \! J_{tS} \! - \! i \frac{\Gamma_{tS}}{2} \! \right) \! \left( \hat{\sigma}_t^\dagger \hat{\sigma}_S + \hat{\sigma}_S^\dagger \hat{\sigma}_t \right),
\end{align}
where $\Delta = \omega_L - \omega_0$ is the laser detuning, and $\delta_{tS}$ captures any additional frequency offset between the target and auxiliary atoms. The coherent and dissipative couplings between the target emitter and the symmetric ring mode are $J_{tS} = \sqrt{N} J_{t1}$ and $\Gamma_{tS} = \sqrt{N} \Gamma_{t1} $, respectively. 
For subwavelength interatomic separation between the ring atoms, \eqnref{eq: nonHermitian_Ham_ring} exhibit a collective dark eigenstate $\ket{\phi_d}$ with a narrowed linewidth $\gamma_d$, whose red-sidebands can be exploted for cooling the target atom.

Figure~\ref{fig: RingCooling}(b) shows the steady state population $n_t$ as a function of the ring radius $R$ and detuning $\delta_{tS}$. In general, smaller ring size $R$ leads to lower values for $n_t$. However, for small rings, the final temperature of the target atoms increase if the ring atoms and the target atoms are resonant ($\delta_{tS}=0$). This occurs because the dipole shift $J_S$ detunes the collective symmetric mode $\ket{S}$ from the target atom thereby suppressing the interference effect responsible for the emergence of the narrowed linewidth $\gamma_d$ [blue curve in Fig.~\ref{fig: RingCooling}(c)]. 
This limitation can be overcome by the following detuning between the target and auxiliary emitters
\begin{equation}
\label{eq: optimal_detuning}
\delta_{tS}^\mathrm{dark} =  J_S - J_{cS} \frac{\Gamma_S- \gamma_c}{\Gamma_{cS}},
\end{equation}
which minimizes the subradiant decay rate and is obtained as the detuning at which the coherent and dissipative parts of the spin Hamiltonian $\hat{H}_S$ commute. 
At small $R$, this detuning reduces the subradiant decay rate, as shown by the orange curve in Fig.~\ref{fig: RingCooling}(c), leading to a heavily reduced final steady state population [black dash-dotted curve in Fig.~\ref{fig: RingCooling}(b)] which coincides with the optimal one, obtained numerically (white dashed line). In this regime, where $\Gamma_S \propto N\gamma_c$ and $\Gamma_{cS} \propto \sqrt{N}\gamma_c$, the strongly subradiant state can be approximated as $\ket{\phi_d} \approx (\sqrt{N} \hat{\sigma}_t^\dagger - \hat{\sigma}_S^\dagger)\ket{0}/\sqrt{N+1}$. The dark mode thus describes two dipole of equal magnitude emitting with opposite phase and population mostly localized at the target atom, thereby resulting in near-perfect destructive interference~\cite{Ritsch_subradiance_9fold}.  
At larger $R$, \eqnref{eq: optimal_detuning} is no longer optimal, because the decay rates of the collective modes are comparable, and it is thus favorable to tune both into resonance.
Notably, $\gamma_d$ and $\bar{n}_t$ remain nearly independent of atom number at fixed $R$ under $\delta_{tS}^\mathrm{dark}$,\footnote{Note that $\delta_{tS}^\mathrm{dark}$ approaches zero for nine-atom rings due to the geometric properties of the Green's function~\cite{Ritsch_subradiance_9fold}. Thus, the steady state phonon number when no detuning is applied (\ie for $\delta_{tS}^\mathrm{dark} = 0$) is minimized for 9-atom rings.} as shown in Fig.~\ref{fig: RingCooling}(d), as $\gamma_d/ \gamma_0 \approx 1- \beta^2/\alpha + O(1/N)$ with $\beta = \gamma_{1t} / \gamma_0$ and $\alpha \geq  \beta^2$ defined by $\Gamma_S \approx \alpha N \gamma_t$~\cite{Ritsch_subradiance_9fold}.
Increasing $N$ beyond moderate values does not significantly improve the subradiant decay rate or steady-state phonon number.

We note that the phonon population of the ring emitters also evolves during cooling of the central atom. However, due to their significantly different trap frequencies, the drive is far off-resonant from any sidebands of the ring atoms. As a result, their cooling and heating rates are strongly suppressed compared to $\mathcal{R}_{tt}^-$, and their phonon number remains essentially unchanged over the timescale required to cool the target emitter.

Our results generalize those of Wang et al. in Ref.~\cite{Wang2023}, which reported a $10\%$ improvement in the target atom’s final temperature only for very specific emitter positions where coherent dipole-dipole interactions vanish. By suitably detuning the target atom and directly driving the subradiant cooling sideband on resonance, our approach removes this positional constraint and achieves over an order-of-magnitude reduction in final temperature compared to an isolated atom.

Finally, cooling an emitter through the collective mode of a ring of emitters bears some analogy with cavity cooling of atoms~\cite{Zippilli2005PRA,Zippilli2005PRL,Cirac1993OptComm,Cirac1995}. However, there are important differences. While cavity cooling exploits an interference effect that suppresses heating sideband transitions, the approach analyzed here leverages dissipative atom-ring coupling to engineer collective sidebands with narrower linewidth. 

\section{Conclusions}\label{sec:Conclusions}

We demonstrated that the modified optical response of dipole-interacting atoms in free space can be harnessed to cool atoms to their motional ground state. Thanks to the narrowed linewidth of collective subradiant states, when the atomic trap frequencies are not identical (as is often the case in experiments), it is possible to cool the entire atomic ensemble to temperatures lower than those achievable for independent (\ie noninteracting) atoms. Notably, we find instances where ground-state cooling is only possible due to light-induced dipole-dipole interactions. We emphasize that our scheme can be implemented using simple laser configurations (for example, by uniformly driving all atoms), requiring only the adjustment of the laser detuning.

These results are obtained from an effective master equation for the motional dynamics of dipole-interacting atoms and are confirmed by extensive numerical simulations of the original model in the Lamb-Dicke approximation. 
The derivation of an effective mechanical model (which holds for generic arrangements of the atoms) in the presence of subradiant states overcomes the limitations of previous approaches~\cite{Palmer2010} by recognizing that atomic center of mass fluctuations broaden the subradiant collective linewidth by $\sim\eta^2\gamma_0$~\cite{Guimond2019,Rusconi2021} and provide insight on the mechanisms that enable enhanced cooling.

Our findings are directly applicable to arrays of alkaline-earth and rare-earth atoms, which exhibit long-wavelength transitions~\cite{olmos_long-range_2013, masson2023dicke} and can be trapped at subwavelength separations using metasurface holographic optical traps~\cite{huang2023metasurface}, optical tweezer arrays~\cite{cooper2018alkaline} or optical lattices~\cite{Rui2020,Derevianko2011optical,Leticia_Strontium}. 
Beyond the case of levitated atoms in free space, these methods could be applied more broadly to cool atoms close to nanofibers~\cite{atoms_nanofiber} or in cavities~\cite{Mishina2014}, or cool levitated objects with embedded quantum emitters such as diamonds with NV centers~\cite{Delord2020}.

Our method relies on the presence of internal atomic resonances for the emergence of collective scattering modes with a modified linewidth. 
It is natural to ask whether these findings can be extended to polarizable objects which do not possess internal resonances. Optically levitated nanoparticles, for instance, have recently attracted significant attention as ideal platform to build sensors with unprecedented sensitivity or to test fundamental physics~\cite{Gonzalez2021levitodynamics,Millen2020}. In such systems, collective scattering modes could potentially be engineered using a multimode cavity or, for larger objects, by exploiting geometric resonances such as Mie resonances in spherical particles. If successful, these ideas could enable ground-state cooling of multiple nanoparticles, opening exciting avenues for exploring out-of-equilibrium dynamics~\cite{nanoparticles_2,Shengyan2020} and non-reciprocal interactions~\cite{nanoparticles_uros,nanoparticles_uros_2,Rudolph2024PRL}, or detection of weak signals~\cite{Gambhir2016,Carney2020,Monteiro2020}.

Finally, we showed that the modified collective response of subwavelength arrays can be used to improve on the control of the motional degrees of freedom of the atoms. Beyond the sideband cooling scenario considered here, we anticipate that these ideas will stimulate further exploration of the correlated dynamics between internal and external degrees of freedom, with the potential to harness these correlations to improve and enhance light–matter interfaces based on subwavelength arrays.

\acknowledgements{O.R.-B. acknowledges support from Fundación Mauricio y Carlota Botton and from Fundació Bancaria “la Caixa” (LCF/BQ/AA18/11680093).
SFY thanks the NSF throughPHY-2207972, the CUA PFC PHY-2317134, and the Q-SEnSE QLCI OMA-2016244.
A. A.-G. acknowledges support from the AFOSR through their Young Investigator Prize (No. 21RT0751), from the NSF through their CAREER Award (No. 2047380), and from the David and Lucile Packard Foundation.
C.C.R. acknowledges support from the European Union’s Horizon Europe program under the Marie Sklodowska Curie Action LIME (Grant No. 101105916). This research was supported in part by the Austrian Science Fund (FWF) [10.55776/COE1 QuantA] and by the grant NSF PHY-2309135 to the Kavli Institute for Theoretical Physics (KITP).}

\appendix

\section{Free-space dipole-dipole interactions}\label{app:Green_Tensor}

The dipole-dipole coupling strength between atoms $i$ and atoms $j$ in free space i given by~\cite{Lehmberg1970a} 
\begin{equation}\label{eq:G_coupling}
    G(\rrop_i -\rrop_i) \equiv  -\frac{3\pi \gamma_0}{k_0} \hat{\dpp}^\dagger \boldsymbol{G}(\rrop_i-\rrop_j,\omega_0)  \hat{\dpp},
\end{equation}
where $\gamma_0=\mu_0\w_0^3|\dpp|^2/(3\pi\hbar c)$ is the single-atom spontaneous emission rate in free space, $\hat{\dpp} = \dpp / |\dpp|$ is the unit vector pointing in the direction of the atomic dipole matrix element $\dpp$. In   $\boldsymbol{G}(\rr)$ is the free space electromagnetic Green's tensor. The general expression for the Green's tensor everywhere except at the origin reads
\begin{equation}\label{eq:green}
\begin{split}
    \boldsymbol{G}(\rr,\w_0) \equiv & \frac{e^{i k_0 r}}{4\pi k_0^2r^3}\Big[(k_0^2r^2+i k_0 r-1)\id\\
    &+(-k_0^2r^2-3i k_0r+3)\frac{\rr\otimes\rr}{r^2}\Big].
\end{split}
\end{equation}
In this article, we define $G(0)=-i\gamma_0/2$, thereby including into the atomic resonance frequency $\omega_0$ the contribution of the single particle Lamb-Shift, proportional to the real part of \eqnref{eq:G_coupling} evaluated at the origin.

\section{Derivation of the effective master equation}\label{app:Derivation_effective_ME}

In this appendix, we derive the effective master equation~(\ref{eq:ME_Motion}) describing the laser cooling of the motional degrees of freedom of the system. The derivation relies on two approximations. First, we assume the internal or spin degrees of freedom to be only weakly excited, which is valid for weak and off-resonant drives. This allows to approximate the optical response of the atomic ensemble by its linear response. Second, that the coupling between internal and external degrees of freedom is smaller than the characteristic timescale of the spin dynamics, which allows for an adiabatic elimination of the spins. 
Below, we derive specific condition for the validity of these two assumptions.
Note that the derived effective master equation includes the effects arising from light-induced dipole-dipole interactions mediated by the vacuum electromagnetic field, and therefore describes collective cooling in dense atomic ensembles. 

\subsection{Displacement transformation in the low spin-excitation regime}

The starting point of our analysis is the master equation for the full system in the Lamb-Dicke approximation, Eq.~(\ref{eq:ME_Lamb-Dicke}).
Under the assumption of weak saturation, $\langle \hat{\sigma}_i^\dagger \hat{\sigma}_i \rangle \ll 1$ $\forall i$, we approximate the spin operators by bosonic operators with $[\hat{\sigma}_i, \hat{\sigma}_j^\dagger] = \delta_{ij}$.
Next, we apply the unitary transformation
\begin{equation}
\label{eq: displacement_transformation}
    \hat{U}_d = \exp \left( \sum_j (s_j^*\hat{\sigma}_j + \beta_{j}^*\hat{b}_{j} -\text{H.c.} ) \right),
\end{equation}
which displaces the bosonic spin and phonon operators as $\hat{\sigma}_i \rightarrow \hat{\sigma}_i + s_i$ and $\hat{b}_i \rightarrow \hat{b}_{i} + \beta_{i}$. 
Equation~\eqref{eq: displacement_transformation} corresponds to a shift of system's degrees of freedom to a new equilibrium position determined by the strength of the external driving laser. 
The displacements can be approximated as a power series in the Lamb-Dicke parameter. To leading order they read 
\begin{equation}
\begin{split}
    s_i =& s_i^{(0)} + \mathcal{O}(\eta^2)\\
    \beta_{i} =& \eta_i \beta_{i}^{(1)} + \mathcal{O}(\eta^3).
\end{split}
\end{equation}
The leading order in the spin-displacement is obtained by solving the linear system
\begin{equation}
\label{eq: displacement_spins}
    \left( -\Delta - i \frac{\gamma_0}{2} \right) s_i^{(0)} + \sum_{j \neq i} \left( J_{ij} - i \frac{\gamma_{ij}}{2} \right) s_j^{(0)} = - \Omega_i.
\end{equation}
Physically, $s_i^{(0)}$ represents the excited-state amplitudes of atom $i$ induced by the driving field. 
For drive strengths much smaller than the laser detuning or the atomic decay rates, $|s_i^{(0)}|\ll1 $ $\forall i$ and the weak-excitation approximation remains valid. The first-order motional displacement reads
\begin{equation}
    \beta_{i}^{(1)} = -\frac{2}{\nu_i} \mathrm{Re} \left\{ \Omega_{i}' s_i^{*(0)} + \sum_{j \neq i} \left( J_{ij}' - i \frac{\gamma_{ij}'}{2} \right) s_i^{* (0)} s_j^{ (0)} \right\},  
\end{equation}
where $\Omega_{i}'$, $J_{ij}'$ and $\gamma_{ij}'$ represent the derivatives of the Rabi frequency and dipole-dipole interactions along the direction of motion. Since all $\beta_{i}^{(1)}$ are real, they simply give rise to a shift of the center-of-mass position of the atoms. That is, only the position of the atoms is displaced up to to second order in $\eta$. 

At this point it is convenient to regroup the different contributions in the master equation into three terms (aside from constant energy shifts which we neglect): 

(i) The first term contains only spin operators and reads
\begin{equation}\label{eq:L_0_app}
	\mathcal{L}_S \muop = - \frac{i}{\hbar}\pare{\Hop'_S\muop - \muop \Hop'^\dag_S}+ \mathcal{D}_S \muop.
\end{equation}
The non-Hermitian Hamiltonian takes the form
\begin{equation}
\label{eq: spin_liouvillian_H}
    \hat{H}'_S = \Hop_S +  \hbar \sum_{i, j\neq i} 2G_{ij}' \mathrm{Re} \{ \eta_i^2 \beta_{i}^{(1)} - \eta_j^2 \beta_{j}^{(1)} \}   \hat{\sigma}_i^\dagger \hat{\sigma}_j,
\end{equation}
where $\Hop_S$ is defined in \eqnref{eq:H_S} and the last term represents and additional correction of order $\eta^2$ arising from the displacement of the atomic position. 
Note that even for $\beta_i^{(1)}=0$, $\Hop_S$ contains a second order corrections in $\eta$ that arises from the zero-point motion of the emitters. 
Formally, this correction is obtained from terms of the form $\hat{b}_{i} \hat{b}_{i} ^\dagger$, which become $\hat{b}_{i}^\dagger \hat{b}_{i}  + 1 $ upon normal ordering (see discussion at the end of Sec.~\ref{sec:Lamb-Dicke_approx}). Finally, the population recycling term reads
\begin{equation}
\label{eq: spin_liouvillian_D}
\begin{split}
    \mathcal{D}_S \muop =&\! \sum_{i,j \neq i} \!\pare{\tilde{\gamma}_{ij}+2\gamma_{ij}' \mathrm{Re} \{ \eta_i^2 \beta_{i}^{(1)} - \eta_j^2 \beta_{j}^{(1)}\}}\! \hat{\sigma}_j \muop \hat{\sigma}_i^\dagger\\
    +&\sum_{i} \left(\gamma_0 +\eta_i^2 \gamma_{ii}''\right)\hat{\sigma}_i \muop \hat{\sigma}_i^\dagger.
\end{split}
\end{equation}

The non-Hermitian Hamiltonian $\Hop'_S$ governs the emission properties of the system~\footnote{Compared to the optical response provided in the main text, $H_S'$ accounts for the additional corrections due to the center-of-mass displacement. For the particular case of perpendicular motion considered in the main text, $G'_{ij}=0$ and the $\Hop'_S=\Hop_S$.}.
Due to its non-Hermitian nature, it diagonalizes in a biorthogonal basis consisting of $N$ right eigenvectors $|\phi_\lambda\rangle = \sum_i \phi_{\lambda,i} \hat{\sigma}_i^\dagger |0\rangle$ and $N$ left eigenvectors $\langle \bar{\phi}_\lambda | = \sum_i \phi_{\lambda,i} \langle 0 | \hat{\sigma}_i $, where $|0\rangle$ is the total ground state of the $N$ bosonized and displaced spins~\cite{Brody2014}. They satisfy
\begin{subequations}
\label{eq: eigenvectors_nonHerm_general}
\begin{align}
    \hat{H}'_S |\phi_\lambda\rangle = \epsilon_\lambda |\phi_\lambda\rangle, \\
    \langle \bar{\phi}_\lambda | \hat{H}'_S = \epsilon_\lambda \langle \bar{\phi}_\lambda |,
\end{align}
\end{subequations}
with complex eigenvalues $\epsilon_\lambda \equiv -\Delta + J_\lambda - i \gamma_\lambda / 2$. The real and imaginary parts, $J_\lambda$ and $\gamma_\lambda$, respectively represent the collective energy shift and decay rate associated with the mode $|\phi_\lambda\rangle$. 
These eigenvectors typically correspond to spin waves delocalized over the full atomic ensemble. For large periodic arrays or emitters in a ring geometry, they are well approximated by states with well-defined quasimomentum~\cite{AsenjoGarcia2017}. Importantly, the eigenmodes define the system’s modified optical response and encode the collective absorption and emission of light characteristic of dipole-coupled emitter ensembles. Notably, the $\eta^2$ correction  to the dissipative dipole-dipole interactions imposes a fundamental lower bound on the achievable decay rates, such that $\min_\lambda(\gamma_\lambda) \sim \eta^2 \gamma_0$. 
As a result, the system cannot support perfectly dark states with vanishing decay~\cite{Guimond2019,Rusconi2021}. This key observation enables the derivation of an effective master equation for the atomic motion, even in large atomic arrays, as will become evident in subsequent subsections.

While different eigenvectors are not orthogonal to one another, they are normalized and satisfy $\langle \phi_\lambda | \phi_\lambda \rangle = \langle \bar{\phi}_\lambda | \bar{\phi}_\lambda \rangle =1$. Noting that the identity matrix can be expressed as $\mathbb{I} = \sum_\lambda | \phi_\lambda \rangle \langle \bar{\phi}_\lambda | / \langle \bar{\phi}_\lambda | \phi_\lambda \rangle$, it follows that
\begin{equation}
\label{eq: biorthogonal_changebasis}
    \hat{\sigma}_i^\dagger |0\rangle = \sum_\lambda \frac{\phi_{\lambda,i}}{\sum_j \phi_{\lambda,j}^2} |\phi_\lambda\rangle \equiv \sum_\lambda c_{\lambda,i} |\phi_\lambda\rangle.
\end{equation}
In the case of degenerate normal modes, one has to select an appropriate basis within the degenerate subspace for which $\langle \bar{\phi}_\lambda | \phi_\lambda \rangle = \sum_i \phi_{\lambda,i}^2 \neq 0$.

(ii) The second contribution involves only motional operators and takes the form
\begin{align}\label{eq:L_M}
	\mathcal{L}_M \muop &= - \frac{i}{\hbar}\big[ \Hop_M , \muop \big]  - \eta^2 \sum_{i,j} \frac{\gamma_{ij}''}{2} s_i^{*(0)} s_j^{(0)} \times \nonumber \\
    & \times \left( 2 \hat{R}_{j} \muop \hat{R}_{i} - \muop \hat{R}_{i}^2 - \hat{R}_{j}^2 \muop + \frac{1}{2} \Big\{ ( \hat{R}_{i} \! - \! \hat{R}_{j})^2 , \muop   \Big\}   \right),
\end{align}
where $\Rop_i = \sqrt{\bar{\nu}/\nu_i}(\bdop_i+\bop_i)$.
The second term in \eqnref{eq:L_M} describes collective recoil heating which is proportional to the dipole products of the dipole moments  of the atoms, $s_i^{*(0)} s_j^{(0)}$. The Hamiltonian $\hat{H}_M$ governing the coherent motional dynamics is given by
\begin{align}
\label{eq: Liovillian_motion_general}
    \frac{\hat{H}_M}{\hbar} &=   \sum_{i} \nu_i \hat{b}_{i}^\dagger \hat{b}_{i}  + \eta^2 \sum_{i}  \mathrm{Re} \{ \Omega_{i}'' s_i^{*(0)} \} \hat{R}_{i}^2 \nonumber \\
    &+ \eta^2 \sum_{i , j\neq i} \frac{J_{ij}''}{2} \mathrm{Re} \{ s_i^{(0)} s_j^{*(0)} \} ( \hat{R}_{i} -  \hat{R}_{j})^ 2.
\end{align}
The second and third term in \eqnref{eq: Liovillian_motion_general} describe respectively a renormalization of the trap frequency due to the drive, and a coupling between different sites due to mechanical forces induced by dipolar interactions $(J_{ij}'')$.

(iii) The third contribution describes the interaction between internal and external degrees of freedom. We define the interaction Liouvillian as 
\begin{equation}
\label{eq: L_int_1D}
\mathcal{L}_\mathrm{int} \muop \equiv \pare{\mathcal{L}_+ +\mathcal{L}_-}\muop
\end{equation}
where $\mathcal{L}_\pm \equiv - (i/\hbar) \left[  \hat{H}_\pm, \muop \right] +  \mathcal{D}_\pm \muop$.
To first order in $\eta$, the interaction reads
\begin{widetext}
\begin{subequations}
\label{eq: interaction_Liouvillian}
\begin{align}
    \frac{\hat{H}_\pm}{ \hbar} &= \sum_{i} \eta_i \left( \Omega_{i}'  \hat{\sigma}_i^\dagger + \Omega_{i}'^{*}   \hat{\sigma}_i \right) \hat{b}_{i}^\pm +  \sum_{i , j\neq i} G_{ij}' \left( \eta_i \hat{b}_{i}^\pm - \eta_j \hat{b}_{j}^\pm \right) \left( \hat{\sigma}_i^\dagger \hat{\sigma}_j + s_j^{(0)} \hat{\sigma}_i^\dagger + s_i^{*(0)} \hat{\sigma}_j \right),\label{eq:H_pm} \\ 
    \mathcal{D}_\pm \muop &=  \sum_{i,j \neq i} \gamma_{ij}^\prime \left[ \eta_i \left( \hat{\sigma}_j \muop \hat{\sigma}_i^\dagger + s_j^{(0)} \muop \hat{\sigma}_i^\dagger + s_i^{*(0)} \hat{\sigma}_j \muop \right) \hat{b}_{i}^\pm - \eta_j \hat{b}_{j}^\pm \left(\hat{\sigma}_j \muop \hat{\sigma}_i^\dagger + s_j^{(0)} \muop \hat{\sigma}_i^\dagger + s_i^{*(0)} \hat{\sigma}_j \muop \right) \right]. 
\end{align}
\end{subequations}
\end{widetext}
Here, we defined $\hat{b}_{i}^- \equiv \hat{b}_{i}$ and $\hat{b}_{i}^+ \equiv \hat{b}_{i}^\dagger$, such that $\mathcal{L}_+ \muop$ and $\mathcal{L}_- \muop$ respectively capture the contributions from motional creation and annihilation operators.

The interaction Hamiltonian admits a clear physical interpretation. The term $\eta_i \Omega_i' \hat{b}_i^\dagger \hat{\sigma}_i^\dagger$ describes the simultaneous creation of a spin and motional excitation in atom $i$ upon absorption of a laser photon. Conversely, $\eta_i \Omega_i' \hat{b}_i \hat{\sigma}_i^\dagger$ corresponds to the excitation of the spin while annihilating a motional excitation, again driven by photon absorption. Similar processes arise from the light-mediated dipole interactions, e.g., $\eta_i J_{ij}' s_j^{(0)} \hat{b}_i^\pm \hat{\sigma}_i^\dagger$, where the role of the laser photon is effectively replaced by a spin excitation at atom $j$, with amplitude proportional to $s_j^{(0)}$. Note that this term vanishes (\ie, $G_{ij}'' = 0$) for atoms moving along any direction perpendicular to the line connecting their center-of-mass positions. These processes are sketched in Fig.~\ref{fig:Cooling_Processes}.

\subsection{Adiabatic elimination of spin degrees of freedom}

To proceed, we establish a hierarchy of timescales relevant to the problem. Specifically, we consider the regime of weak optomechanical coupling defined by the condition 
\begin{equation}\label{eq:Weak_Coupling_Condition}
     \eta_i \Omega_i , \eta_i J_{ij}' |s_j^{(0)}|  ,\eta_i \gamma_{ij}' |s_j^{(0)}|\ll \nu_i, \gamma_\lambda \quad \forall i,j,\lambda.
\end{equation}
Equation~\eqref{eq:Weak_Coupling_Condition} is typically satisfied in the Lamb-Dicke regime ($\eta_i \ll 1$) and under weak excitation limit ($\Omega_i \lesssim \gamma_0$ such that $|s_i^{(0)}| \ll 1$). Under these conditions, the rapidly evolving internal dynamics can be adiabatically eliminated, and an effective equation for the atomic motional dynamics can be obtained by the following procedure.

First, we move to the interaction picture with respect to the free motional Hamiltonian $\hbar \sum_{i} \nu_i \hat{b}_{i}^\dagger \hat{b}_{i}$, under which the motional operators transform as $\hat{b}_{i} \rightarrow \hat{b}_{i} e^{-i \nu_i t}$. While the spin Liouvillian $\mathcal{L}_S \muop$ remains unchanged, the interaction Liouvillian becomes time-dependent,
\begin{equation}\label{eq:L_int_1D_interaction_pic}
\begin{split}
    \mathcal{L}_\mathrm{int} (t) \muop =& \spare{\mathcal{L}_{+}(t) + \mathcal{L}_{-}(t)} \muop\\
    =&\sum_{i} \pare{e^{i \nu_i t} \mathcal{L}_{+,i} + e^{-i \nu_i t} \mathcal{L}_{-,i}} \muop,
\end{split}
\end{equation}
where we defined $\mathcal{L}_{\pm} = \sum_i \mathcal{L}_{\pm,i}$, with $\mathcal{L}_{\pm,i}$ containing only motional operators of emitter $i$.

To derive an effective equation of motion for the atomic motion, we need to define the relevant subspace where the effective dynamics takes place.
Accordingly, we define a superoperators $\mathcal{P}$ that projects the dynamics into the subspace where the internal degrees of freedom are in their displaced ground state $\ket{0}$, 
$\mathcal{P}\muop = \Tr_S[\muop]\otimes \ketbra{0}{0}\equiv \rhop\otimes \ketbra{0}{0}$.
We then apply the projector operator formalism for open quantum systems~\cite{Javanainen1984,Stenholm1986,cirac1992laser,gonzalezballestero2023tutorial} and derive the Nakajima-Zwanzig equation. 
Under the time scale separation defined by \eqnref{eq:Weak_Coupling_Condition}, we approximate the Nakajima-Zwanzig equation to second order in the interaction Liouvillian \eqref{eq: L_int_1D} as
\begin{equation}
\label{eq: adiab_elim_formula_priorRWA}
    \frac{d}{dt} \mathcal{P}\muop =  \mathcal{L}_M(t) \mathcal{P}\muop +  \mathcal{L}_\mathrm{int} (t)\!\! \int_0^\infty \!\!\!d\tau\, e^{\tau \mathcal{L}_S} \mathcal{L}_\mathrm{int}(t-\tau) \mathcal{P}\muop .
\end{equation}
Notably, the second term in Eq.~(\ref{eq: adiab_elim_formula_priorRWA}) is of second order  in the interaction Liouvillian $\mathcal{L}_\mathrm{int}(t)$. Using the decomposition in \eqnref{eq:L_int_1D_interaction_pic}, we write \eqnref{eq: adiab_elim_formula_priorRWA} as the sum of four terms. Two of these terms are obtain as the double action of $\mathcal{L}_+$ or of $\mathcal{L}_-$. In the interaction picture they contain sum of operators that oscillate as  $\propto e^{i \tau (\mathcal{L}_S \pm i \nu_i)} e^{\pm i (\nu_i + \nu_j) t}$, and can therefore be neglected under the rotating wave approximation when \eqnref{eq:Weak_Coupling_Condition} holds.
The remaining two terms, instead, contains operators that oscillate as $e^{i \tau (\mathcal{L}_S \pm i \nu_i)} e^{i t (\nu_i- \nu_j)}$ and must therefore be kept.
Eliminating the fast oscillating terms from \eqnref{eq: adiab_elim_formula_priorRWA}, and taking the trace over the internal degrees of freedom we obtain
\begin{align}
\label{eq: adiab_elim_formula_RWA}
    \frac{d}{dt} \hat{\rho} = & \, \mathcal{L}_M \hat{\rho} \Big|_\mathrm{RWA} + \sum_{l = \pm} \sum_{i,j} e^{i l (\nu_i - \nu_j)t} \, \mathrm{Tr}_S \Bigg\{ \mathcal{L}_{-l,j} \nonumber \\
    & \times \int_0^\infty d\tau \, e^{\tau (\mathcal{L}_S - i l \nu_i)} \mathcal{L}_{l,i} \hat{\rho} \otimes |0\rangle \langle 0| \Bigg\}.
\end{align}

Let us now evaluate the second term in Eq.~(\ref{eq: adiab_elim_formula_RWA}). To illustrate the derivation in a concise and transparent manner, we restrict our attention to the contribution in $\mathcal{L}_{\pm,i}$ that is proportional to the first derivative of the drive; the remaining terms can be treated analogously. Focusing on this component, one readily finds
\begin{align}
    \mathcal{L}_{\pm,i} \hat{\rho} \otimes |0\rangle \langle 0| = 
    - i \eta_i \Big( & \Omega_{i}' \hat{b}_{i}^\pm \hat{\rho} \otimes \hat{\sigma}_i^\dagger |0\rangle \langle 0| \nonumber \\
    & - \Omega_{i}'^{*} \hat{\rho} \hat{b}_{i}^\pm \otimes |0\rangle \langle 0| \hat{\sigma}_i \Big).
\end{align}
We now apply the time-dependent superoperator $e^{\tau (\mathcal{L}_S - i l \nu_i)}$, which evolves the state under the fast spin dynamics and the free evolution of the atomic motion. To facilitate the calculation, we express the operators $\hat{\sigma}_i^\dagger |0\rangle \langle 0|$ and $|0\rangle \langle 0| \hat{\sigma}_i$ in terms of the eigenstates of the spin Liouvillian $\mathcal{L}_S$ introduced in~\eref{eq: eigenvectors_nonHerm_general},
\begin{subequations}
\begin{align}
    \mathcal{L}_S  |\phi_\lambda\rangle \langle 0|  &= -i \epsilon_\lambda |\phi_\lambda\rangle \langle 0|, \\
    \mathcal{L}_S  |0\rangle \langle \phi_\lambda| &= i \epsilon_\lambda^* |0\rangle \langle \phi_\lambda|.
\end{align}
\end{subequations}
The action of the superoperator $e^{\tau (\mathcal{L}_S \mp i \nu_i)}$ is straightforward. Transforming back to the local spin basis $\hat{\sigma}_i^\dagger |0\rangle$, we  obtain
\begin{align}\label{eq:Application_L_int_once}
    &e^{\tau (\mathcal{L}_S  \mp i \nu_i )}  \mathcal{L}_{\pm,i} \hat{\rho} \otimes |0 \rangle \langle 0| = \nonumber \\
    &\quad - i \eta_i  \sum_\lambda \sum_j e^{-i \tau (\epsilon_\lambda \pm\nu_i)} \Omega_{i}' \, c_{\lambda,ij} \, \hat{b}_{i}^\pm \hat{\rho} \otimes \hat{\sigma}_j^\dagger |0 \rangle \langle 0| \nonumber \\
    &\quad + i \eta_i \sum_{\lambda} \sum_j e^{i \tau (\epsilon_\lambda^* \mp \nu_i)} \Omega_{i}'^* \, c_{\lambda,ij}^* \, \hat{\rho} \hat{b}_{i}^\pm \otimes |0\rangle \langle 0| \hat{\sigma}_j ,
\end{align}
where we have defined the coefficients $c_{\lambda,ij} = \phi_{\lambda,i} \phi_{\lambda,j} / \sum_m \phi_{\lambda,m}^2$. While the spin-motion interaction induced by the driving field is local (\ie it acts individually on each emitter), the delocalized character of the spin eigenstates introduces non-locality into the dynamics. This effect is evinced by the appearance of an additional atomic index $j$, and ultimately gives rise to nonlocal interactions between the motional degrees of freedom.

The temporal integral in \eqnref{eq: adiab_elim_formula_RWA} can now be simply calculated and reads
\begin{align}
\label{eq: timeintegral}
    \int_0^\infty d\tau\, e^{-i \tau (\epsilon_\lambda \pm \nu_i)} = \frac{-i}{ \epsilon_\lambda \pm \nu_i} \equiv E_{\lambda i\pm}.
\end{align}
Finally, we apply the superoperator $\mathcal{L}_{-l,j}$ to the result of \eqnref{eq:Application_L_int_once} after the integration in \eqnref{eq: timeintegral}, and trace over the spin degrees of freedom. This yields nonzero contributions only from terms of the form $\mathrm{Tr}_S \{ |0\rangle \langle 0 | \}$ and $\mathrm{Tr}_S \{ \hat{\sigma}_i^\dagger |0\rangle \langle 0 | \hat{\sigma}_i \}$. Upon additionally performing the rotating wave approximation (RWA) on the motional Liouvillian $\mathcal{L}_M(t) \hat{\rho}$ in the interaction picture, Eq.~(\ref{eq: adiab_elim_formula_RWA}) reduces to the final master equation governing the atomic motion given in \eqnref{eq:ME_Motion}.

\subsection{Effective master equation and rate equations for the atomic motion}

The explicit form of the effective master equation for the atomic motion is presented in Eq.~(\ref{eq:ME_Motion}) of the main text. The first term, proportional to $V_{ij}$, captures coherent interactions between the motional degrees of freedom, mediated by both the light field and the collective spin excitations. The second ($\propto \mathcal{R}_{ij}^-$) and third ($\propto \mathcal{R}_{ij}^+$) terms are Lindblad dissipators that respectively account for collective motional cooling and heating processes—that is, the correlated loss or gain of motional quanta.

The expectation value of any motional operator $\hat{O}$ can be computed from the reduced density matrix as $\langle \hat{O} \rangle = \mathrm{Tr} \{ \hat{O} \hat{\rho} \}$. Using the effective master equation~(\ref{eq:ME_Motion}), we can derive the equations of motion for the phonon populations $\langle \hat{b}_i^\dagger \hat{b}_i \rangle$ and their coherences $\langle \hat{b}_i^\dagger \hat{b}_j \rangle$:
\begin{align}
\label{eq: EOM_pops_general}
    \frac{d}{dt} \langle \hat{b}_i^\dagger \hat{b}_j \rangle &= i \left( \nu_i - \nu_j \right) \langle \hat{b}_i^\dagger \hat{b}_j \rangle  + \mathcal{R}_{ij}^{+} \nonumber \\
    &+ \sum_{m=1}^N \left(i V_{mi} - \frac{\mathcal{R}_{mi}^{-}}{2} + \frac{\mathcal{R}_{im}^{+}}{2} \right) \langle \hat{b}_m^\dagger \hat{b}_j \rangle \nonumber \\ 
    &+ \sum_{m=1}^N \left( -i V_{jm} - \frac{\mathcal{R}_{jm}^{-}}{2} + \frac{\mathcal{R}_{mj}^{+}}{2} \right) \langle \hat{b}_i^\dagger \hat{b}_m \rangle.
\end{align} 
The quadratic structure of the effective master equation ensures that the expectation values of second-order operators do not couple to higher- or lower-order moments. This closed set of equations allows for efficient computation of both the dynamics and steady-state values of the phonon numbers. In particular, the steady state is obtained by solving the linear system of $N^2$ equations obtained by setting $d\langle \hat{b}_i^\dagger \hat{b}_j \rangle / dt = 0$.

The coherent and dissipative coupling rates in Eqs.~(\ref{eq:ME_Motion}) and~(\ref{eq: EOM_pops_general}) arise from two main contributions. The first originates from recoil-induced diffusion, captured by the terms proportional to $\eta^2$ in Eq.~(\ref{eq:L_M}). The second contribution results from the interference between laser- and dipole-induced spin-motion couplings, given in Eq.~(\ref{eq: L_int_1D}), which are proportional to $\Omega_j'$ and $G_{ij}'$, respectively.

In Eq.~(\ref{eq: dissipative_interactions_perp}) of the main text, we present the cooling and heating transition rates for atoms moving along a direction perpendicular to the line
connecting their equilibrium positions, such that $G_{ij}' = 0$ $\forall i,j$ and the only contribution to cooling comes from the sideband excitation of the laser. Note that the steady state coherences in Eq.~(\ref{eq: dissipative_interactions_perp}) are equal to the spin displacements, $\avg{\sigma_j}_\text{ss} \equiv s_j^{(0)}$ and $\avg{\sigma_i^\dag}_\text{ss} \equiv s_i^{*(0)}$. In this configuration, the coherent energy shifts ($V_{ii}$) and coupling rates ($V_{ij}$ with $i \neq j$) are given by
\begin{subequations}
\begin{align}
    V_{ii} &=  \eta_i^2 \sum_{j \neq i} \mathrm{Re} \Big\{ \left( 2J_{ij}'' -i \gamma_{ij}'' \right) s_i^{*(0)} s_j^{(0)}  \Big\}   \nonumber \\
    &+ 2 \eta_i^2 \mathrm{Re} \{ \Omega_{i}'' s_i^{*(0)} \} + \Delta_{ii}^{-} + \Delta_{ii}^{+}   \label{eq:V_jj_gen},  \\
    V_{ij} &= \Delta_{ij}^{-} + \Delta_{ji}^{+} - 2 \eta_i \eta_j J_{ij}'' \mathrm{Re} \{ s_i^{(0)} s_j^{*(0)} \}, \label{eq:V_ij_gen}\\
    \Delta_{ij}^{\pm} &= -\frac{1}{2} \eta_i \eta_j  \Omega_j' \Omega_i'^{*} \sum_\lambda \left(  \frac{c_{\lambda,ij}}{\epsilon_\lambda \pm \nu_{j}} + \frac{c_{\lambda,ij}^*}{\epsilon_\lambda^* \pm \nu_{i}} \right).\label{eq: SI_rates_coh}
\end{align}
\end{subequations}

In the general case where atoms can move along the direction connecting their equilibrium positions, the coupling rates receive additional contributions from the first-order spatial derivatives of the dipole-dipole interactions, $G_{ij}'$. The resulting cooling and heating transition rates are given by
\begin{align}\label{eq:Rate_General}
\mathcal{R}_{ij}^{\pm} & = \eta_i \eta_j \sum_\lambda \left( A_{\lambda j} B_{\lambda i}^* E_{\lambda j\pm} + A_{\lambda i}^* B_{\lambda j}  E_{\lambda i\pm}^* \right) \nonumber \\
&-  \eta_i \eta_j \gamma_{ij}'' s_j^{(0)} s_i^{*(0)},
\end{align}
where the last term corresponds to recoil heating. The first two terms, where we defined the coeffiencients
\begin{align}
    A_{\lambda i} =& - i \Omega_i' c_{\lambda,i} - i \sum_{j \neq i} \left( J_{ij}' -i \frac{\gamma_{ij}'}{2}  \right) \left( c_{\lambda,i} s_j^{(0)} + c_{\lambda,j} s_i^{(0)} \right) , \nonumber \\
    B_{\lambda i} =& - i \Omega_i' \phi_{\lambda,i}^* - i \sum_{j \neq i} J_{ij}' \left( \phi_{\lambda,i}^* s_j^{(0)} + \phi_{\lambda,j}^* s_i^{(0)} \right)\nonumber \\
    & - \sum_{j \neq i} \frac{\gamma_{ij}'}{2}  \left( \phi_{\lambda,i}^* s_j^{(0)} - \phi_{\lambda,j}^* s_i^{(0)} \right),
\end{align}
encapsulate the interference between the contribution of laser- and dipole-induced spin-motion interactions to first order in $\eta$, which are respectively depicted in Fig.~\ref{fig:Cooling_Processes}(a) and (b).  

The coherent energy shifts and couplings retain the same structure as in Eqs.~(\ref{eq:V_jj_gen}) and (\ref{eq:V_ij_gen}), with the terms $\Delta_{ij}^\pm$ replaced by the modified contributions $\tilde{\Delta}_{ij}^\pm$, given by
\begin{equation}
    \tilde{\Delta}_{ij}^{\pm}  = \frac{1}{2} \eta_i \eta_j \sum_\lambda \left( - i A_{\lambda j} B_{\lambda i}^* E_{\lambda j\pm} + i A_{\lambda i}^* B_{\lambda j}  E_{\lambda i\pm}^* \right) .
\end{equation}

Finally, it is instructive to compare the relative strengths of the laser- and dipole-induced contributions to the dissipative and coherent motional interactions. As seen in the expression for $A_{\lambda i}$, the laser-induced terms scale with $\Omega_i'$, while the dipole-induced terms scale with $G_{ij}' s_j^{(0)}$. As we discuss in detail for the two-atom case in Appendix~\ref{app:Generalizations}, the dipole-induced contributions are strongly suppressed in the regime of large trap frequencies $\nu \gg \gamma_0$, where $s_j^{(0)} \sim \mathcal{O}(\Omega_j / \nu)$.

\section{Generalized formalism for motion along multiple directions} \label{app:Formalism_Multiple_Directions}

In this appendix, we generalize the formalism to account for atomic motion in all three spatial dimensions. 

\subsection{Lamb-Dicke approximation}

Let us first derive the Lamb-Dicke master equation for the full system in the presence of motion along several directions.
Following the same procedure as in Sec.~\ref{sec:Lamb-Dicke_approx}, we expand the total master equation~(\ref{eq:ME_Total}) to second order in the Lamb-Dicke parameter $\eta$, and obtain
\begin{equation}\label{eq:ME_Lamb-Dicke_3D}
	\partial_t \muop = \pare{\mathcal{L}_0 + \eta \mathcal{L}^{(1)}_\text{int} + \eta^2 \mathcal{L}^{(2)}_\text{int}} \muop.
\end{equation}
To avoid heavy notation, we are using the same labels for the different contribution in \eqnref{eq:ME_Lamb-Dicke_3D} as the one we used in the main text, although the terms here include motion along different directions.

The first term in Eq.~(\ref{eq:ME_Lamb-Dicke_3D}) describes the dynamics in the absence of spin-motion coupling, and reads 
\begin{equation}\label{eq:L_0_3D}
\begin{split}
	\mathcal{L}_0 \muop =& - \frac{i}{\hbar}\pare{\Hop_0\muop - \muop \Hop_0^\dag}+ \sum_{i,j} \gamma_{ij} \hat{\sigma}_j \muop \hat{\sigma}_i^\dag.
\end{split}
\end{equation}
The non-Hermitian Hamiltonian $\hat{H}_0$ takes the form
\begin{align}\label{eq:H_0_3D}
    \hat{H}_0 &= \hbar\sum_{j} \sum_\alpha \nu_{j\alpha} \bdop_{j\alpha} \bop_{j\alpha} + \hbar \sum_j \left( \Omega_j \hat{\sigma}_j^\dagger + \Omega_j^* \hat{\sigma}_j \right) \nonumber \\
    & - \hbar\Delta\sum_j \hat{\sigma}^\dag_j\hat{\sigma}_j + \hbar \sum_{i,j} G_{ij} \hat{\sigma}^\dag_i\hat{\sigma}_j,  
\end{align}
where $\nu_{j\alpha}$ is the trap frequency at site $j$ along the spatial direction $\alpha \in \{ x,y ,z\}$. We defined the bosonic annihilation operator $\hat{b}_{j \alpha}$ for a motional quantum at site $j$ along direction $\alpha$, as well as the corresponding dimensionless position operators $\hat{R}_{j \alpha}\equiv \sqrt{\bar{\nu}/\nu_{j \alpha}}(\bdop_{j \alpha}+\bop_{j \alpha})$. Here, the average trap frequency is defined as $\bar{\nu} \equiv \sum_j \sum_\alpha \nu_{j \alpha} / 3N$, from which we introduce the averaged Lamb-Dicke parameter $\eta \equiv \sqrt{ \nu_R / \bar{\nu}}$. We further define the site- and direction-dependent Lamb-Dicke parameter as $\eta_{j \alpha} = \sqrt{ \nu_R / \nu_{j\alpha}}$.

The second term in Eq.~(\ref{eq:ME_Lamb-Dicke_3D}) corresponds to the spin-motional coupling to first order in the Lamb-Dicke parameter. It reads
\begin{align}\label{eq:L_1_3D}
    \mathcal{L}_\text{int}^{(1)} \muop &= -\frac{i}{\hbar} \pare{\Hop_1\muop - \muop\Hop_1^\dag}  \nonumber \\
    &+ \sum_{i\neq j} \sum_\alpha \gamma_{ij}^{(\alpha)} \hat{\sigma}_j \left( \muop  \hat{R}_{i \alpha} - \hat{R}_{j \alpha}  \muop \right) \hat{\sigma}_i^\dagger,
\end{align}
where the non-Hermitian Hamiltonian takes the form
\begin{align}\label{eq:H_1_Lamb-Dicke_3D}
    \hat{H}_{1} &=  \hbar \sum_j \sum_\alpha \left( \Omega_j^{(\alpha)}  \hat{\sigma}_j^\dagger +\hc \right)  \hat{R}_{j \alpha} \nonumber \\
    &+ \hbar \sum_{ij} \sum_\alpha \!G_{ij}^{(\alpha)} \left( \hat{R}_{i \alpha} - \hat{R}_{j\alpha} \right) \hat{\sigma}_i^\dagger \hat{\sigma}_j.
\end{align}
Here, we define the first-order derivatives of the Rabi frequency and the dipole-dipole interaction along the spatial direction $\alpha$ as $\W_j^{(\alpha)} \equiv k_0^{-1}\partial \W(\rr)/\partial r_{\alpha}|_{ \rr_{0j}}$, and $G_{ij}^{(\alpha)} \equiv J_{ij}^{(\alpha)} - i \gamma_{ij}^{(\alpha)}/2 \equiv k_0^{-1} \partial G (\mathbf{r})/\partial r_\alpha |_{\mathbf{r}_{0i}-\mathbf{r}_{0j}}$.

The last term in \eqnref{eq:ME_Lamb-Dicke_3D} contains interaction terms to second order in the Lamb-Dicke parameter, and reads
\begin{equation}\label{eq:L_2_3D}
\begin{split}
	\mathcal{L}_\text{int}^{(2)}\muop =& - \frac{i}{\hbar}\pare{\Hop_2 \muop - \muop \Hop_2^\dag}  \\
	+&  \sum_{i, j} \sum_{\alpha, \beta} \frac{\gamma_{ij}^{(\alpha \beta)}}{2} \hat{\sigma}_j \left( \muop\Rop_{i \alpha} \Rop_{i \beta} \! + \!  \Rop_{j\alpha} \Rop_{j \beta}\muop \right) \hat{\sigma}_i^\dagger \\
    -&  \sum_{i, j} \sum_{\alpha, \beta} \gamma_{ij}^{(\alpha \beta)} \hat{\sigma}_j  \Rop_{j \alpha} \muop \Rop_{i \beta}  \hat{\sigma}_i^\dagger,
\end{split}
\end{equation}
where we defined the non-hermitian Hamiltonian
\begin{equation}\label{eq:H_2_3D}
\begin{split}
	\Hop_2 &= \frac{\hbar}{2} \sum_j \sum_{\alpha,\beta} \left( \Omega_j^{(\alpha\beta)}  \hat{\sigma}_j^\dagger + \hc \right)  \hat{R}_{j\alpha} \hat{R}_{j\beta} \\ &+ \frac{\hbar}{2}\sum_{i,j} \sum_{\alpha,\beta} G_{ij}^{(\alpha\beta)} \big(\hat{R}_{i \alpha} - \hat{R}_{j \alpha}\big) \big(\hat{R}_{i \beta} - \hat{R}_{j \beta}\big)  \hat{\sigma}_i^\dagger \hat{\sigma}_j.
\end{split}
\end{equation}
The second-order derivatives along the spatial directions $\alpha$ and $\beta$ are defined as $\W_j^{(\alpha \beta)} \equiv k_0^{-2}\partial^2 \W(\rr)/\partial r_{\alpha} \partial r_{\beta}|_{ \rr_{0j}}$ and $G_{ij}^{(\alpha \beta)} \equiv J_{ij}^{(\alpha \beta)} - i \gamma_{ij}^{(\alpha \beta)}/2 \equiv k_0^{-2} \partial^2 G (\mathbf{r})/\partial r_\alpha \partial r_\beta |_{\mathbf{r}_{0i}-\mathbf{r}_{0j}}$. 

\subsection{Effective master equation for the atomic motion}

Following a similar procedure as in Appendix~\ref{app:Derivation_effective_ME}, we derive the effective master equation for the atomic motion that accounts for dynamics along multiple spatial directions
\begin{equation}
\label{eq:ME_Motion_directions}
\begin{split}
    \frac{d}{dt} \hat{\rho} =& -i\Big[\sum_{i,j}\sum_{\alpha, \beta}V_{i \alpha j \beta}e^{\ii (\nu_{i\alpha} - \nu_{j\beta}) t}\bdop_{i\alpha}\bop_{j\beta},\hat{\rho}\Big]\\
     +&  \sum_{i,j}\sum_{\alpha,\beta}\! \mathcal{R}_{i \alpha j \beta}^- e^{ \ii (\nu_{i\alpha} - \nu_{j\beta}) t} \Big( \hat{b}_{j\beta} \hat{\rho} \hat{b}_{i\alpha}^\dagger \!- \! \frac{1}{2} \{\hat{b}_{i\alpha}^{\dagger} \hat{b}_{j\beta} ,\hat{\rho}\} \Big) \\
     +&  \sum_{i,j}\sum_{\alpha,\beta}\! \mathcal{R}_{i \alpha j \beta}^+ e^{ -\ii (\nu_{i\alpha} - \nu_{j\beta}) t} \Big( \hat{b}_{j\beta}^\dagger \hat{\rho} \hat{b}_{i\alpha} \!- \! \frac{1}{2} \{\hat{b}_{i\alpha} \hat{b}_{j\beta}^{\dagger},\hat{\rho}\} \Big) .
\end{split}
\end{equation} 
The cooling ($\hat{R}_{i \alpha j \beta}^-$) and heating ($\hat{R}_{i \alpha j \beta}^+$) transition rates, as well as the coherent interactions $V_{i \alpha j \beta}$, originate from the laser- and dipole-induced spin-motion interactions in Eq.~(\ref{eq:L_1_3D}), as well as from recoil heating effects described by Eq.~(\ref{eq:L_2_3D}). For simplicity, we present only the contributions arising from the laser-induced spin-motion coupling and recoil heating, which give rise to the dissipative transition rates 
\begin{align}
\label{eq: SI_rates_dissipative_directions}
\mathcal{R}_{i \alpha j \beta}^{\pm} = &-i \eta_{i \alpha} \eta_{j \beta} \Omega_j^{(\beta)} \Omega_i^{(\alpha)*} \sum_\lambda \left(  \frac{c_{\lambda,ij}}{\epsilon_\lambda \pm \nu_{j \beta}}  - \frac{c_{\lambda,ij}^*}{\epsilon_\lambda^* \pm \nu_{i \alpha}} \right)\nonumber\\
 -&  \eta_{i \alpha} \eta_{j \beta} \gamma_{ij}^{(\alpha \beta)} s_j^{(0)} s_i^{*(0)}, 
\end{align}
as well as the coherent interaction terms
\begin{align}
\label{eq: SI_rates_coherent_directions}
    V_{i\alpha i\beta} &=  \eta_{i \alpha} \eta_{i \beta} \sum_{j \neq i} \mathrm{Re} \Big\{ \left( 2 J_{ij}^{(\alpha \beta)} -i\gamma_{ij}^{(\alpha \beta)} \right)  s_i^{*(0)} s_j^{(0)} \Big\} \nonumber \\
    &+ 2 \eta_{i \alpha} \eta_{i \beta} \mathrm{Re} \Big\{ \Omega_{i}^{(\alpha \beta)} s_i^{*(0)} \Big\} + \Delta_{i\alpha i \beta}^- + \Delta_{i \beta i \alpha}^+, \nonumber \\
    V_{i\alpha j \beta} &=  - 2 \eta_{i\alpha} \eta_{j\beta} J_{ij}^{(\alpha \beta)} \mathrm{Re} \Big\{ s_i^{*(0)} s_j^{(0)} \Big\} + \Delta_{i\alpha j \beta}^- + \Delta_{j \beta i \alpha}^+,\nonumber \\
    \Delta_{i\alpha j \beta}^{\pm} &= -\frac{1}{2} \eta_{i \alpha} \eta_{j \beta}  \Omega_j^{(\beta)} \Omega_i^{(\alpha)*} \sum_\lambda \! \left( \! \frac{c_{k,ij}}{\epsilon_\lambda \pm \nu_{j \beta}} + \frac{c_{\lambda,ij}^*}{\epsilon_\lambda^* \pm \nu_{i \alpha}} \! \right).
\end{align}
For the case of motion along multiple directions, the spin displacements $s_i^{(0)}$ remain given by Eq.~(\ref{eq: displacement_spins}), while the collective spin eigenmodes $|\phi_\lambda\rangle$ (as well as their overlaps $c_{\lambda,ij}$ and eigenvalues $\epsilon_\lambda$) are obtained by diagonalizing the non-Hermitian spin Hamiltonian in Eq.~(\ref{eq: spin_liouvillian_H}), with the modified dipole-dipole interaction $\tilde{G}_{ij} = G_{ij} + \sum_\alpha G_{ij}^{(\alpha \alpha)} (\eta_{i\alpha}^2 + \eta_{j \alpha}^2)/2$.

\section{Numerical simulation of the cooling dynamics}\label{app:Numerics}

The results presented in this work are obtained either by solving the adiabatically eliminated equations~\eqref{eq:ME_Motion} for the atomic motion only, or by numerically solving the master equation~(\ref{eq:ME_Lamb-Dicke}) for the full model taking into account both the spin and motional degrees of freedom.  

In order to numerically solve Eq.~(\ref{eq:ME_Lamb-Dicke}), we employ the computational framework QuantumOptics.jl ~\cite{kramer2018quantumoptics}, which allows for a straightforward implementation of the equations on a computer. Nevertheless, the dimension of the Hilbert space of spin and motional degrees of freedom increases exponentially with the number of atoms $N$, namely $\mathrm{dim}(\mathcal{H})=2^N \times N_\mathrm{cutoff}^N$ where $N_\mathrm{cutoff}$ is the maximal number of motional excitations per atom. For two atoms, we solve the equations for $N_\mathrm{cutoff} = 2$. For larger systems we solve Eq.~(\ref{eq:ME_Lamb-Dicke}) by truncating the Hilbert dimension down to sizes that can be efficiently simulated~\cite{Suresh2022}. We consider that at most one atomic spin can be excited at a time, a valid approximation in the weak driving regime and for far off-resonant laser drives. Similarly, we also assume that the atoms share at most one phonon or motional quantum, which limits the validity of the numerical treatment to situations where the atoms are ground-state cooled.
Under these approximations, the dimension of the Hilbert space is reduced to $\mathrm{dim}(\mathcal{H})= (N+1)^2$. Consequently, the total density matrix $\muop$ in Eq.~(\ref{eq:ME_Lamb-Dicke}) has dimension $(N+1)^2 \times (N+1)^2$ and we are able to simulate its time evolution up to $N=7$ atoms. For that, we convert $\muop$ into a vector $\Vec{\mu}$ defined on a Fock-Liouville space~\cite{manzano2020short}. Then, Eq.~(\ref{eq:ME_Lamb-Dicke}) can be recast as $d\Vec{\mu} /dt= \mathcal{L} \Vec{\mu}$, where $\mathcal{L}$ is the Liouville superoperator describing both the effect of the Hamiltonian and the population recycling term. The state of the system over time is finally obtained via matrix exponentiation as $\Vec{\mu}(t) = \exp (\mathcal{L} t) \Vec{\mu}(0)$, where $\Vec{\mu}(0)$ is the initial density matrix describing both the spin and motional degrees of freedom. Finally, the cooling rate is extracted by fitting an exponential decaying phonon number over time.

\begin{figure}
    \centering
    \includegraphics[width=\columnwidth]{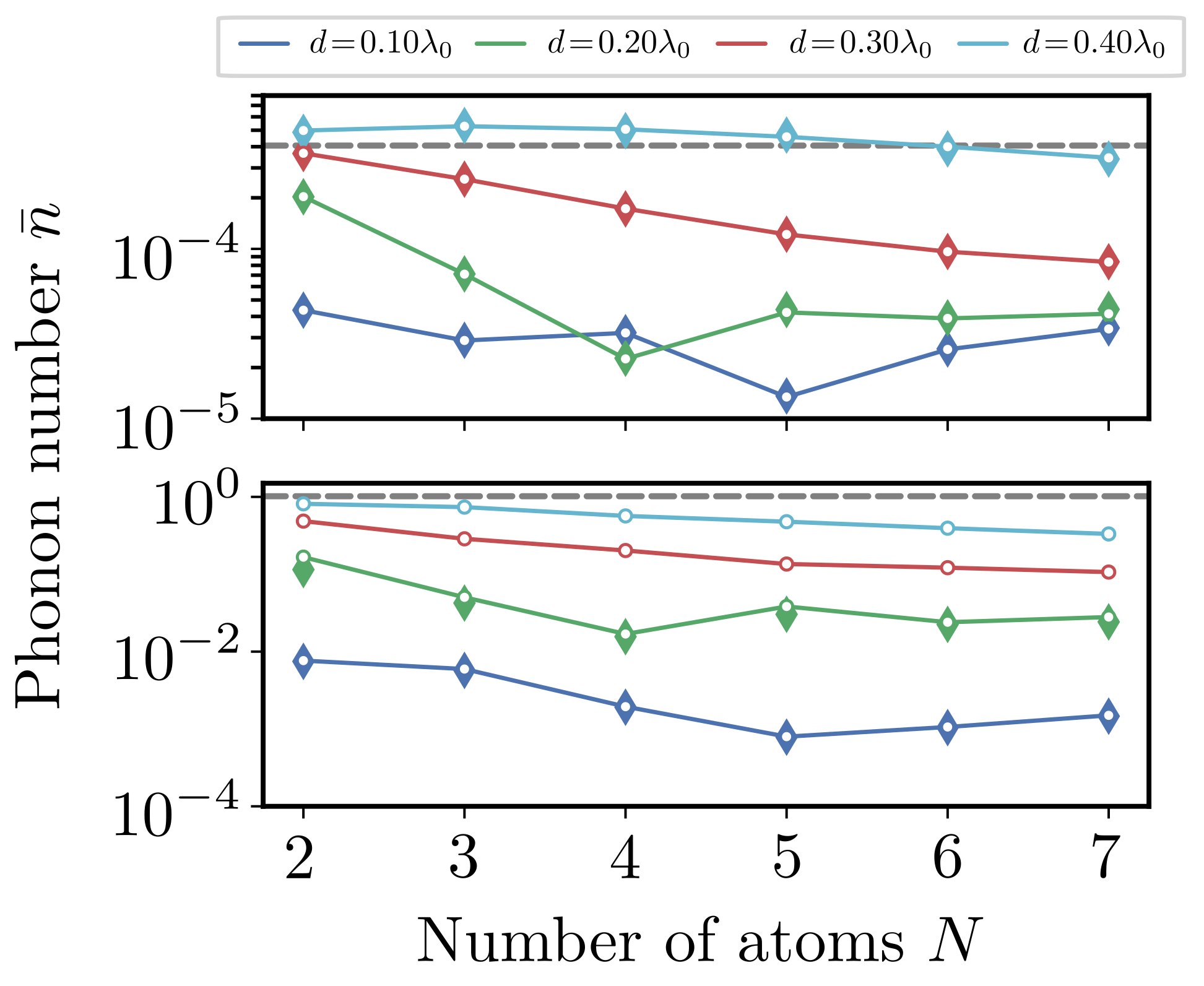}
    \caption{Steady-state population $\bar{n}$ for a one-dimensional array of up to $N=7$ atoms moving perpendicular to the array direction. Different colors represent different lattice spacing as specified by the legend. Top and bottom panel refer to the case where $\bar{\nu}=20\gamma_0$ and $\bar{\nu}=0.25\gamma_0$ respectively. We consider $\nu_j = \bar{\nu} + \delta \nu (j - \lfloor N/2 \rfloor -1)$ in both panels, such that the system is in the \emph{array cooling} regime. The empty circles represent the result obtain from the effective master equation \eqref{eq:ME_Motion}, while filled diamond are results from the numerical simulation of the Lamb-Dicke master equataion \eqref{eq:ME_Lamb-Dicke}. Due to the large steady-state phonon populations for $d = 0.3\lambda_0$ and $d = 0.4\lambda_0$ at $\bar{\nu} = 0.25\gamma_0$, numerical simulations of \eqnref{eq:ME_Lamb-Dicke} require cutoffs $N_\text{cutoff}$ beyond our computational capabilities and were therefore not performed. }
    \label{fig:Theoy_Simulation}
\end{figure}

In \figref{fig:Theoy_Simulation}, we compare the steady-state phonon population for a chain of $N$ atoms obtained from two approaches: (i) solving the rate equations derived from the effective master equation~\eqref{eq:ME_Motion}, and (ii) numerically simulating the full Lamb-Dicke master equation~\eqref{eq:ME_Lamb-Dicke} using the method described in this section. 
The excellent agreement between the two methods validates the accuracy of the effective model.

\section{Effects of polarization, different direction of motion and coupling to other motional directions}\label{app:Generalizations}

We generalize the case considered in the main text by discussing the role of atomic polarization, the case of atom moving along the direction of separation, and the effects of coupling to other directions. 

\subsection{Dependence on the transition dipole moment }

\begin{figure*}
    \centering   \includegraphics[width = \textwidth]{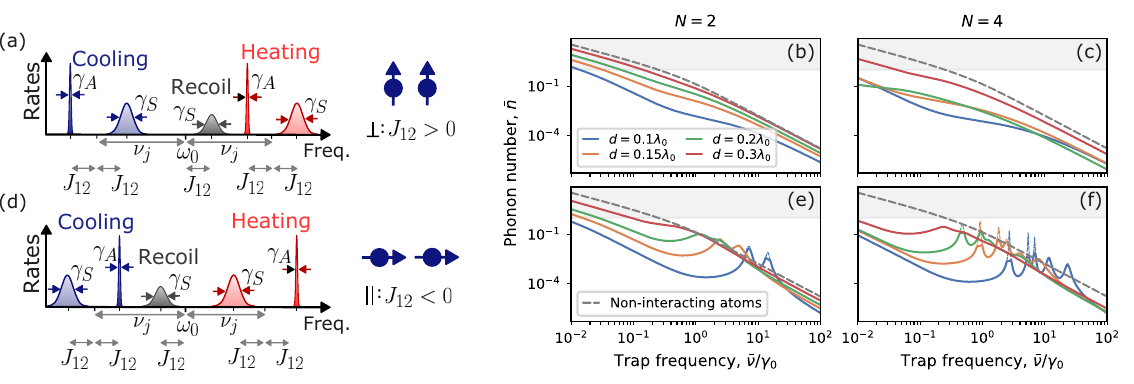}
    \caption{Dependence on atomic polarization. (a) Sideband structure for atoms polarized perpendicular to the axis of the chain. The antisymmetric cooling sideband, symmetric heating sidebands, and the recoil Lorentzian cannot overlap as $d$ decreases. As a result, the phonon number $\bar{n}$ remains below that of noninteracting emitters for all $\bar{\nu}$, both for (b) chains of two atoms and (c) chains of four atoms. (d) Sideband structure for atoms polarized parallel to the axis of the chain. The antisymmetric cooling sideband can become resonant with symmetric heating sidebands and the recoil Lorentzian. The steady state phonon number $\bar{n}$ for (e) chains of two atoms and (f) chains of four atoms can become larger than that of noninteracting atoms for specific combinations of $\bar{\nu}$ and $J_{12}$. The solid lines correspond to a drive on resonance with the cooling sideband of the collective spin mode that leads to a smallest phonon number, whereas the dashed lines correspond to a drive on resonance with the cooling sideband of the darkest spin mode. Notably, when the cooling sideband of the darkest spin mode becomes resonant with either a heating sideband of another spin mode or the recoil Lorentzian, cooling can be enhanced by driving a spin mode with a larger linewidth for which this coalescence does not occur. The results are obtained for $\eta=0.02$ and $\nu_j = \bar{\nu} + \delta \nu (j - \lfloor N/2 \rfloor -1)$ with $\bar{\nu} = 20 \gamma_0$ and $\delta \nu = 5 \times 10^{-4} \gamma_0$. }
    \label{fig: SI_polarization}
\end{figure*}

In this section, we discuss the dependence of the steady state phonon number attained in the array cooling regime on the direction of atomic polarization. For that, it is instructive to first consider the case of two atoms located at a distance $d$ along the $x$-axis and moving along the $z$-direction. The optical response of the system features: (a) two cooling sidebands associated to the antisymmetric and symmetric collective spin modes, respectively centered at $\Delta = - \nu_1 - J_{12}$ and $\Delta = - \nu_1 + J_{12}$; (b) two heating sidebands associated to the antisymmetric and symmetric collective spin modes, respectively centered at $\Delta =  \nu_1 - J_{12}$ and $\Delta =  \nu_1 + J_{12}$; (c) a heating contribution from the recoil or diffusive motion in the form a Lorentzian centered at $\Delta = J_{12}$. As we have done throughout this work, we consider the drive to be on resonance with the antisymmetric cooling sideband. Then, the symmetric heating sideband and the contribution from recoil are respectively detuned from the driving field by $\delta_{hs} = 2 (\nu_1 + J_{12}) $ and $\delta_r = \nu_1 + 2 J_{12} $. There exist two different scenarios depending on the sign of the coherent dipole-dipole interaction $J_{12}$ at small interatomic distances, which in turn depends on the atomic polarization [Fig.~\ref{fig: SI_polarization}(a) and (d)].

For polarization perpendicular to the array and the drive (\ie along $\mathbf{y})$, we obtain $J_{12}>0$ for $d < 0.7 \lambda_0$. Since both the trap frequency $\nu_j$ and $J_{12}$ are positive, the detunings $\delta_{hs}$ and $\delta_r$ cannot become zero. In other words, the symmetric heating sideband and the heating contribution arising from recoil can never be on resonance with the antisymmetric cooling sideband, as illustrated in Fig.~\ref{fig: SI_polarization}(a). As a result, the steady state photon number in the array cooling regime and at small distances $d$ is always smaller than that of noninteracting atoms, as shown in Fig.~\ref{fig: SI_polarization}(b). Note that this is the configuration considered in the main text.
 
For atoms polarized along the direction of the atomic chain (\ie along $\mathbf{x})$, we obtain $J_{12}<0$ for $d < 0.43 \lambda_0$. As illustrated in Fig.~\ref{fig: SI_polarization}(d), it is now possible for the antisymmetric cooling sideband to become resonant with either the symmetric heating sideband or the recoil Lorentzian. This respectively occurs when $\nu_1 = - J_{12}$ and $\nu_1 = - 2 J_{12}$ and leads to increased steady-state phonon numbers as shown in Fig.~\ref{fig: SI_polarization}(e). In particular, the phonon number can become larger than that of noninteracting atoms, such that the cooling advantage arising from the narrow collective resonance is lost. In general, this effect can be minimized by driving on resonance with the cooling sideband associated to another spin mode (in this case, the symmetric cooling sideband), or by separating the atoms by distances such that the different resonances do not overlap for the specific trap frequency considered.

This phenomenon also occurs for larger atom numbers. In Fig.~\ref{fig: SI_polarization}(c) and (f), we show an example for a chain of four atoms. For perpendicular polarization, the most subradiant spin mode is also associated to the most negative energy shift, and efficient cooling is attained for all spacings and trap frequencies. For polarization along the atomic chain, that is no longer the case and cooling via the most subradiant spin mode leads to phonon numbers larger than that of noninteracting atoms for certain combinations of $\bar{\nu}$ and $d$ that lead to coalescence of sidebands. Again, it is generally possible to minimize this extra heating by driving on resonance with the cooling sideband associated to another collective spin mode for which this coalescence does not occur [see the reduced temperature for the solid lines as opposed to the dashed lines in Fig.~\ref{fig: SI_polarization}(f)]. Alternatively, there always exists a spacing $d$ such that the array can be cooled to temperatures lower than that of noninteracting atoms independently of $\bar{\nu}$.

\subsection{Dependence on the axis of atomic motion}

In the main text, we demonstrate enhanced cooling of interacting chains of atoms located on the $x$ axis and moving along the perpendicular direction, that is, along the $z$ axis. Then, the first-order spin-motion interaction in $\eta$, given by \eqnref{eq:H_pm}, receives contributions solely from the laser drive, and is proportional to $\Omega_j'$. In this section, we extend the analysis to atomic motion along the chain, \ie along the $x$ axis, such that the spin-motion coupling acquires an additional contribution from dipole-dipole interactions, proportional to $G_{ij}'$. 

\begin{figure}
    \centering   \includegraphics[width=\columnwidth]{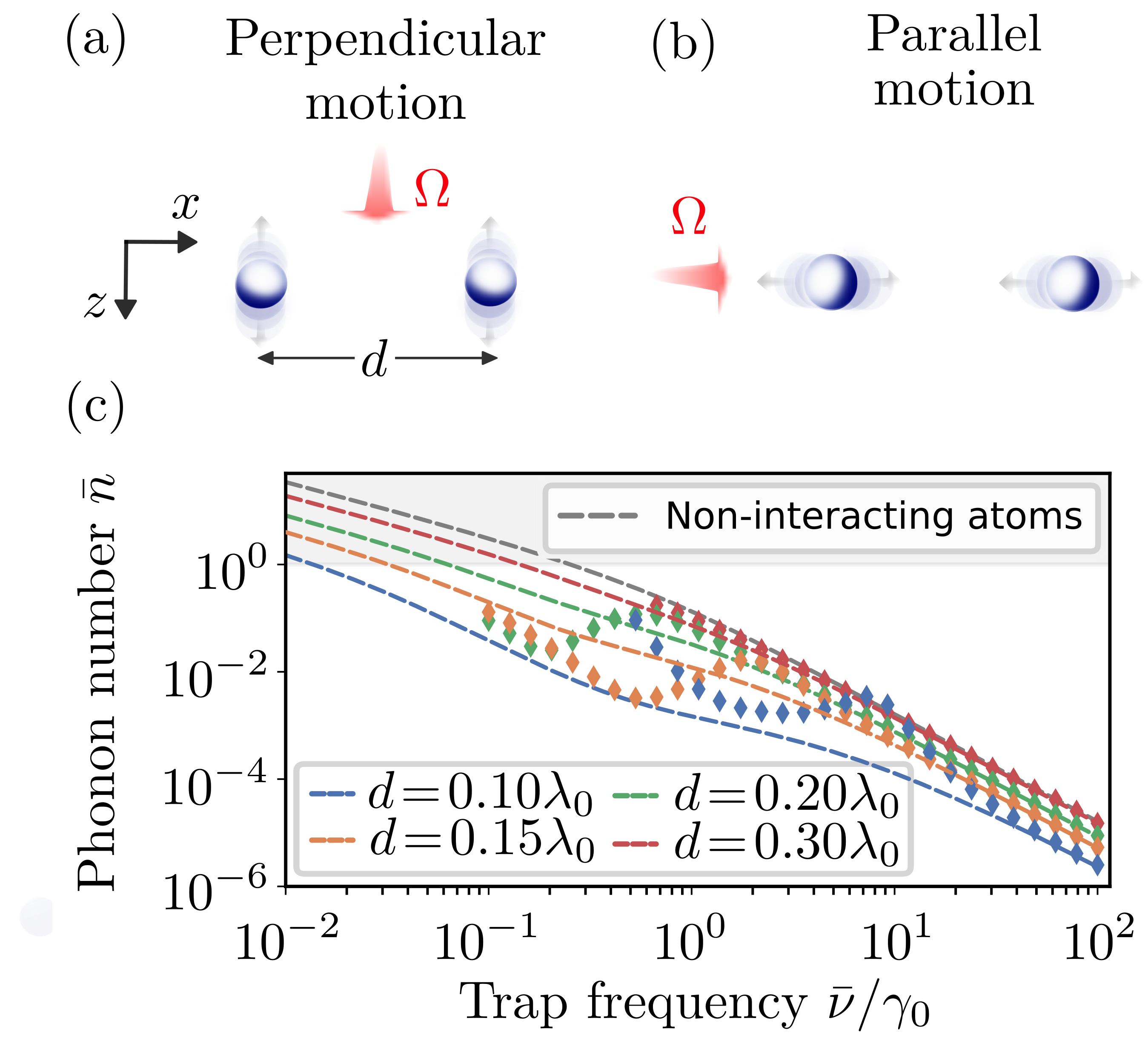}
    \caption{(a) Two trapped atoms at a distance $d$ along the $x$ axis move in the perpendicular direction ($z$ axis). The drive is applied along the direction of motion, such that $\Omega_j = \Omega e^{i k_0 z_j} \equiv \Omega$. (b) Two trapped atoms at a distance $d$ along the $x$-axis move in the same direction (parallel motion). The drive is applied along the direction of motion, such that $\Omega_j = \Omega e^{i k_0 x_j}$. (c) Steady-state phonon number per atom $\bar{n}$ for perpendicular (dashed lines) and parallel (markers) motion, respectively obtained by solving the effective master equation~(\textcolor{blue}{1}) and the full master equation~(\ref{eq:ME_Lamb-Dicke}) including spin and motion. We consider $\eta = 0.02$ and $\delta \nu =  10^{-3} \gamma_0$. 
    }
    \label{fig: SI_motiondirection}
\end{figure}

We consider the case of two atoms placed at a distance $d$. 
For drive along the $x$ direction, the Rabi frequency and it's derivative read $\Omega_j = \Omega e^{i k_0 x_j}$ and $\Omega_j' = i \Omega_j$, as illustrated in Fig.~\ref{fig: SI_motiondirection}(b). For drive on resonance with the antisymmetric cooling sideband $\Delta \approx - \bar{\nu} - J_{12}$, the spin displacements take the form
\begin{equation}
    s_1^{(0)} = -\Omega \frac{ \bar{\nu} + J_{12} \left(1 - e^{i k_0 d} \right) - i \left( \gamma_0 + e^{i k_0 d} \gamma_{12} \right)/2 }{\left( \bar{\nu} - i\gamma_A/2  \right) \left( \bar{\nu} + 2 J_{12} -i \gamma_S /2  \right)},
\end{equation}
and similarly for $s_2^{(0)}$. For $\bar{\nu} \gg |J_{12}|,|J_{12}'|,\gamma_{12},\gamma_{12}'$, the displacement can be approximated to $s_{j}^{(0)} \approx - \Omega_j/\bar{\nu}$. Then, the strength of the spin-motion coupling arising from light-induced dipole-dipole interactions is much smaller than the contribution from the driving field, \ie $\eta_j |J_{12}'| |s_j^{(0)}| \approx \eta_j |\Omega_j| |J_{12}'|/\bar{\nu} \ll \eta_j |\Omega_j| $, and the steady state phonon number $\bar{n}$ is identical for parallel and perpendicular motion. This is shown in Fig.~\ref{fig: SI_motiondirection}(c), where we compare $\bar{n}$ for parallel motion obtained by solving the full master equation~(\ref{eq:ME_Lamb-Dicke}) including the spin and motional degrees of freedom (markers) with $\bar{n}$ for perpendicular motion obtained by solving the effective master equation~(\textcolor{blue}{1}). For $|J_{12}'| \gtrsim \bar{\nu}$, the dipole-dipole contribution to the spin-motion interaction can become dominant, and the steady state phonon number for parallel and out-of-plane motion can differ. As shown in Fig.~\ref{fig: SI_motiondirection}(c), this divergence occurs at larger $\bar{\nu}$ for smaller atomic separations $d$ as $|J_{12}'|$ increases. Notably, enhanced cooling is still found for a wide range of trap frequencies $\bar{\nu}$ and distances $d$.

\subsection{Coupling to other directions of motion}

In this section, we show that enhanced cooling along a single spatial direction remains achievable even when atomic motion occurs in multiple directions. This conclusion is supported both analytically—via the effective master equation in Eq.~(\ref{eq:ME_Motion_directions})—and numerically, through full master equation simulations that account for both spin and motional degrees of freedom.

We consider the same setup as in the main text: The atoms are arranged along the $x$-axis and driven by a classical plane-wave field propagating along $\mathbf{z}$, with drive parameters $\Omega_j = \Omega$ and $\Omega_j^{(\alpha)}=i\Omega \delta_{\alpha z}$. In this configuration the laser directly cools only the motion along the $z$-axis. 
The dissipative and coherent couplings between motion along different directions arise from two different processes. The first is recoil heating, and vanishes since $G_{ij}^{\alpha \beta} = 0$ for all pairs of directions $\alpha \neq \beta$ [see Eq.~(\ref{eq: SI_rates_dissipative_directions})]. The second is the interference between laser- and dipole-induced spin-motion coupling to first-order in $\eta$, given by Eq.~(\ref{eq:L_1_3D}). Notably, $\Omega_j^{(\alpha)}$ and $G_{ij}^{(\beta)}$ are only nonzero for $\alpha=z$ and $\beta=x$, respectively. 
Thus, motion along $\mathbf{z}$ only couples to motion along $\mathbf{x}$ (through terms of the form $\Omega_j^{(z)} G_{ij}^{(x)}$), while motion along $\mathbf{y}$ remains decoupled. For trap frequencies $\nu \gg \gamma_0$, this residual coupling is much smaller than the cooling rates along $\mathbf{z}$, $\mathcal{R}_{iziz}^-$, and thus it does not impact the cooling dynamics. When $\nu \lesssim \gamma_0$, such couplings can still be suppressed under the rotating wave approximation by applying slightly different trap frequencies along $\mathbf{x}$ and $\mathbf{z}$. In this regime, motional degrees of freedom along different directions evolve independently, and enhanced cooling along $\mathbf{z}$ remains possible under the array cooling conditions discussed in the main text. On the other hand, motion along $\mathbf{x}$ and $\mathbf{y}$ can still experience heating, either from recoil (for both directions) or from dipole-induced spin-motion coupling (for $\mathbf{x}$). However, since these heating rates are typically much smaller than the cooling rate along $\mathbf{z}$, they do not lead to appreciable heating during the cooling process.

\begin{figure}[t]
    \centering
    \includegraphics[width =\columnwidth]{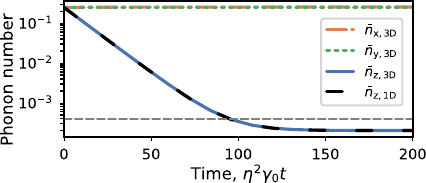}
    \caption{Two atoms separated by $d = 0.2\lambda_0$ along the $x$ axis, polarized along the $y$-axis and subject to a three-dimensional harmonic confinement with trapping frequencies $\nu_{1x}=\nu_{1y}=\nu_{1z} = 20\gamma_0$ for the first atom and $\nu_{1x}=\nu_{1y}=\nu_{1z} = 20.01\gamma_0$ for the second atom. A drive with Rabi frequency $\Omega = 0.1 \gamma_0$ is applied along the $z$ direction, and we further consider $\eta = 0.02$. We plot the average photon number over time along all three directions of motion, $\bar{n}_\mathrm{\alpha,3D}$ with $\alpha \in \{ x,y,z \}$. The phonon number along the $z$ direction, $\bar{n}_\mathrm{z,3D}$, is identical to the case where motion along the $x$ and $y$ directions is neglected, $\bar{n}_\mathrm{z,1D}$. }
    \label{fig: SI_3D_motion}
\end{figure}

We verify this analytical understanding by simulating the cooling dynamics of two atoms moving along all three directions, as shown in Fig.~\ref{fig: SI_3D_motion}. The atoms are located at a distance $d=0.2\lambda$ apart along the $x$-axis, are polarized along $\mathbf{y}$ and are driven by a plane wave field propagating along $\mathbf{z}$. In the array cooling regime and for homogeneous trap frequencies along all three directions, the motion along $\mathbf{z}$ is cooled at the same rate and to the same phonon number as in the absence of motion along the other directions (see blue solid and black dashed curves). Additionally, no significant heating is observed for motion along $\mathbf{x}$ and $\mathbf{y}$ during the cooling process (orange and green curves). 

\section{Numerical optimization of the cooling rate}
\label{App: Optimization_Cooling_Rates}

\begin{figure*}
    \centering   \includegraphics[width=2\columnwidth]{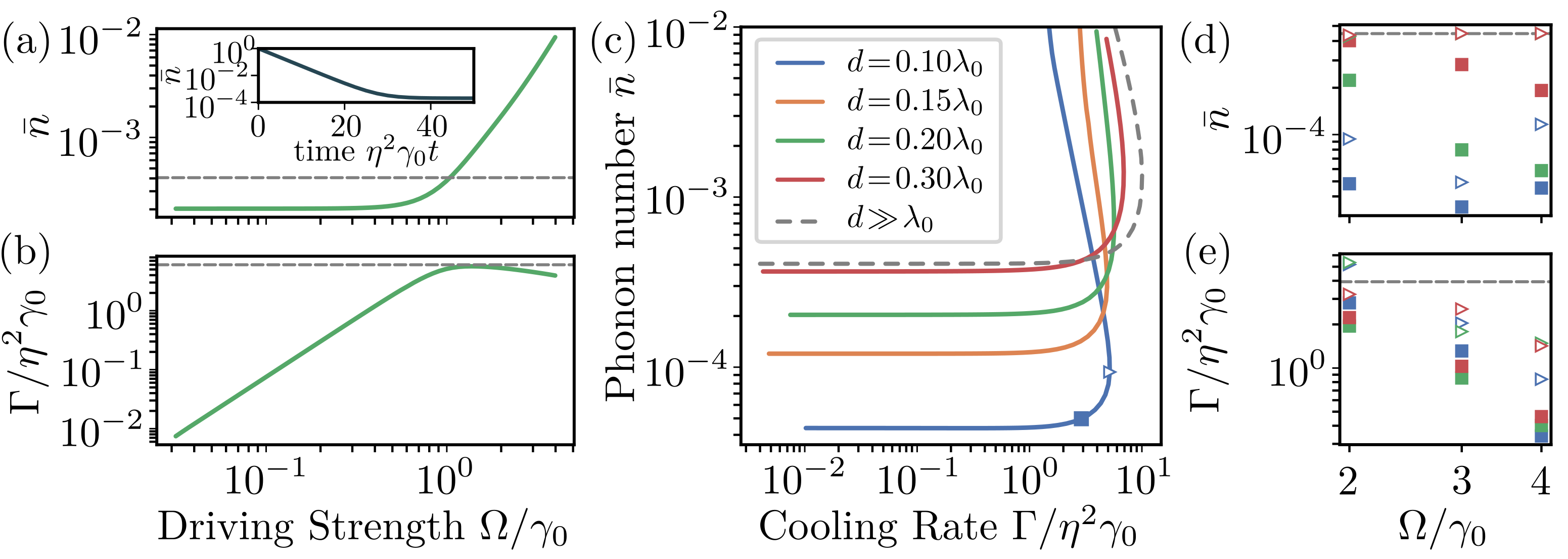}
    \caption{ (a) Steady-state phonon number $\bar{n}$ and (b) cooling rate $\Gamma$ as a function of drive strength $\Omega$ for two atoms separated by a distance $d=0.2\lambda_0$ along the $x$ axis, polarized along the $y$ axis and moving along the $z$ axis. The inset shows an instance of the exponentially decaying phonon number over time for $\Omega = 0.2\gamma_0$, from which the cooling rate is fitted. (c) Steady state phonon number versus cooling rate for various spacings. Each point corresponds to a different value of $\Omega$. The square marker corresponds to the critical cooling rate $\Gamma_c$, obtained at the drive strength that results in a steady state phonon number $10 \%$ larger than the optimal one. The triangular marker corresponds to the alternative critical cooling rate $\Gamma_c'$, obtained at the drive that results in a largest cooling rate while mantaining a steady state phonon number smaller than $1.1 \bar{n}_\textit{ind}$. (d-e) Critical cooling rates $\Gamma_c$ and $\Gamma_c'$ and their corresponding steady state phonon numbers for chains of $N$ atoms and various spacings. In all panels, we consider $\eta = 0.02$ and $\nu_j = \bar{\nu} + \delta \nu (j - \lfloor N/2 \rfloor -1)$ with $\bar{\nu} = 20 \gamma_0$ and $\delta \nu = 5 \times 10^{-3} \gamma_0$. }
    \label{fig: SI_coolingrates_1}
\end{figure*}

In this appendix, we describe how we obtain the critical cooling rates presented in \figref{fig: N_atoms_gradient}(b,d).

Let us consider as an example two atoms in the array cooling regime as defined in \eqnref{eq:array_cooling_condition} and where the subradiant cooling sideband are resolved (\ie  $\gamma_A\ll\bar{\nu}$). The steady state phonon number $\bar{n}$ and the cooling rate $\Gamma$ at which it is reached depend on the drive strength. For sufficiently small drive strength, such that the system remains in the array cooling regime $\Omega^2 \ll \delta \nu \gamma_A / 2 \eta^2 $ and the spin-motion coupling is weak $\Omega \ll \gamma_A/\eta$, the phonon number exponentially decays to its steady state value at a rate $\Gamma = \mathcal{R}^- - \mathcal{R}^+ \approx \mathcal{R}^- \approx 2 \eta^2 \Omega^2/\gamma_A$, as shown in Fig.~\ref{fig: SI_coolingrates_1}(a) and (b) for $\bar{\nu}=20\gamma_0$. As the drive strength increases, either the array cooling condition or the weak spin-motion coupling condition breaks down, and the effective master equation resulting in \eqnref{eq:ME_Motion} stops being valid. In that case, $\bar{n}$ and $\Gamma$ can only be obtained by numerically solving the full system including spin and motional degrees of freedom. More precisely, $\Gamma$ is obtained by fitting an exponential decay to the phonon number over time, as shown in the inset of Fig.~\ref{fig: SI_coolingrates_1}(a). We find that, as soon as one of the two conditions is violated, the phonon number starts to increase with increasing $\Omega$ and the cooling rate eventually decreases. To better understand this process, we plot $\bar{n}$ as a function of cooling rate in Fig.~\ref{fig: SI_coolingrates_1}(c). That is, each data point in this plot corresponds to a different value of $\Omega$. In the main text, we defined the critical cooling rate $\Gamma_c$ as the one obtained at the driving strength $\Omega$ that leads to a steady state phonon number $10\%$ larger than the optimal one. For two atoms at a distance $d=0.1\lambda_0$, this corresponds to the squared marker in Fig.~\ref{fig: SI_coolingrates_1}(c). However, this condition is more restrictive for lattices with darker spin modes than attain lower phonon numbers. In other words, we could drive the two atoms at a distance $d=0.1\lambda_0$ stronger and thereby attain a larger cooling rate while still reaching steady state phonon numbers smaller than in the case of independent atoms, $\bar{n} < \bar{n}_\textit{ind}$. This motivates us to introduce an alternative definition for the critical cooling rate $\Gamma_c'$ as the largest possible rate that can be reached such that the phonon number remains smaller than $1.1 \bar{n}_\textit{ind}$. As an example, $\Gamma_c'$ and its corresponding $\bar{n}'$ are shown by the triangular marker in Fig.~\ref{fig: SI_coolingrates_1}(c) for $d=0.1\lambda_0$. In Figs.~\ref{fig: SI_coolingrates_1}(d) and (e), we plot $\bar{n}$ and $\Gamma_{c}$ for chains of $N$ atoms with various spacing $d$ for $\bar{\nu}=20\gamma_0$. Again, squared and triangular markers denote the two different definitions for the critical cooling rate. In all cases, $\Gamma_c' \gtrsim 2 \Gamma_c$, showing that the critical cooling rates presented in the main text can be sped up by driving the system stronger at the expense of reaching larger steady state phonon numbers $\bar{n}'$. This suggests the following protocol for efficient cooling: first, the system is cooled at $\Gamma'$ to a phonon number close to $\bar{n}'$; then, the drive is dynamically reduced until the lowest possible phonon number $\bar{n}$ is eventually reached.

\section{Cooling $N$ atoms in different geometries}\label{app:Geometries}

\begin{figure}
    \centering   \includegraphics[width=0.9\columnwidth]{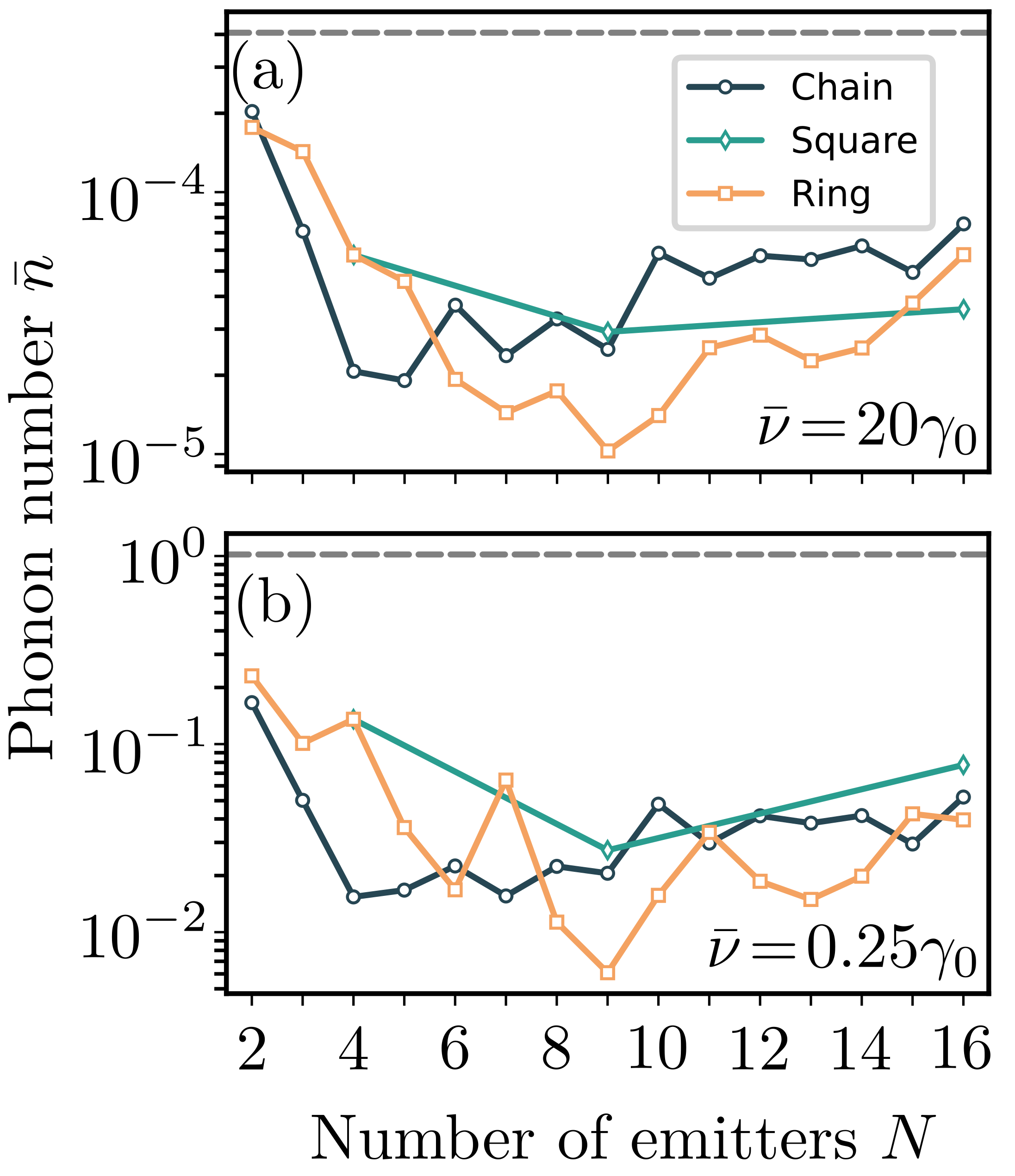}
    \caption{Steady-state phononon number $\bar{n}$ for arrays of different geometries as a function of atom number $N$, for (a) $\bar{\nu} = 20 \gamma_0$ and (b) $\bar{\nu} = 0.25 \gamma_0$. We consider a linear ordered chain, a square two-dimensional lattice and a ring of atoms. The atomic chain is placed along the $x$ axis and the atoms are polarized along the $y$ direction. In all cases, the atoms move along the $z$-direction. The squared array and the ring are placed in the $xy$ plane and the atoms are circularly polarized in the $xy$ plane. For the atomic chain and the squared array, we consider a lattice spacing $d=0.2 \lambda_0$, while the ring has a radius such that neighboring atoms are at a distance $d=0.2 \lambda_0$. The results are obtained for $\eta = 0.02$ and all atoms have different trap frequencies in increments of $\delta \nu = 5 \times 10^{-4} \gamma_0$, such that the system is in the array cooling regime. }
    \label{fig: SI_geometries}
\end{figure}

The results presented in the main text for ordered chains of atoms can be readily extended to other geometries, such as two-dimensional squared arrays or rings of atoms. To exemplify this versatility, in Fig.~\ref{fig: SI_geometries} we show the steady state phonon number as a function of atom number $N$ for $\bar{\nu}=20\gamma_0$ and $\bar{\nu} = 0.25 \gamma_0$ and for atoms arranged in chains, rings, and squares. For the system to be in the array cooling regime, we consider different atoms to have different trap frequencies. For the chain and the ring of atoms, we apply a gradient profile $\nu_j = \bar{\nu} + \delta \nu \left( j - \lfloor N/2 \rfloor -1 \right)$, where $\lfloor N/2 \rfloor$ denotes the integer value of $N/2$ rounded down. For the two-dimensional array, the gradient is applied in both directions of the lattice, such that all atoms have different trap frequencies in increments of $\delta \nu$. We find enhanced cooling to phonon numbers lower than that of non-interacting atoms for all geometries up to $16$ atoms.

\bibliography{main}
\end{document}